\documentclass{amsart}

\usepackage{amssymb}

\newcommand{\rset}{\mathbb{R}}

\numberwithin{equation}{section}

\newtheorem{theorem}{Theorem}[section]

\newtheorem{lemma}[theorem]{Lemma}

\theoremstyle{definition}

\newtheorem{remark}[theorem]{Remark}

\begin{document}

\markboth{Anders Szepessy}{Stochastic molecular dynamics derived from quantum dynamics}

%
%
%
%

\title{LANGEVIN  MOLECULAR DYNAMICS DERIVED FROM EHRENFEST DYNAMICS} 

\keywords{Langevin equation; Ehrenfest dynamics; quantum classical molecular dynamics; 
Gibbs distribution; heat bath; Brownian particle; ab initio molecular dynamics, Mori-Zwanzig theory.}

\subjclass[2000]{82C31, 60H10, 82C10.}

\author{ANDERS SZEPESSY
}

\address{Department of Mathematics,
  Kungl. Tekniska H\"ogskolan,
  100 44 Stockholm,
  Sweden\\
szepessy@kth.se}

\maketitle

%

\begin{abstract}
Stochastic Langevin molecular dynamics for nuclei is derived from the Ehrenfest Hamiltonian system (also called quantum classical molecular dynamics)
in a Kac-Zwanzig setting, with the initial data for the electrons stochastically perturbed from the  ground state
and the ratio, $M$, of nuclei and electron mass tending to infinity.
The Ehrenfest nuclei dynamics is  approximated by the Langevin dynamics
with accuracy $o(M^{-1/2})$ on bounded time intervals and by $o(1)$ on unbounded time intervals,
which makes 
the  small $\mathcal{O}(M^{-1/2})$  friction  and $o(M^{-1/2})$ diffusion terms visible.
The initial electron probability distribution is a Gibbs density at low temperture, 
motivated by a stability and consistency argument.
The diffusion and friction coefficients in the Langevin equation
satisfy  the Einstein's fluctuation-dissipation relation.
\end{abstract}

\section{Introduction to Ab Initio Molecular Dynamics}\label{intro} 
One method to simulate molecular motion is to use quantum classical molecular dynamics (QCMD),
also called {\it Ehrenfest dynamics}, 
where the nuclear positions $X_n:[0,\infty)\rightarrow \rset^3,\ n=1,\ldots ,N$
and the electron wave function $\bar\psi:[0,\infty)\times \rset^{3J}\rightarrow \mathbb{C}$ solve
the  Hamiltonian system %
\begin{eqnarray}
M\ddot X_n^t &=& -\langle \bar\psi^t, {\partial_{X_n}H(X^t)}
\bar\psi^t\rangle ,\label{X_eq}\\
i \frac{d}{dt}\bar\psi^t &=& H(X^t)\bar\psi^t,\label{psi_eq}
\end{eqnarray}
see Refs.   \cite{marx_hutter},   \cite{tully_2}.  
The wave function belongs to a Hilbert space  with inner product $\langle \cdot, \cdot\rangle$ 
and the operator $H(X)$ is 
self-adjoint in that Hilbert space. In computational chemistry the operator
$H$, the electron Hamiltonian, is  precisely determined by the sum of 
kinetic energy of the electrons and the Coulomb interaction between nuclei and electrons; the Hilbert space
is then a subset of $L^2(\rset^{3J})$ with symmetry conditions based on the Pauli exclusion principle for electrons, see Refs.   \cite{lieb_matter},   \cite{lebris_hand}.
Therefore the complex valued $L^2$ inner product
\[
\langle \bar\psi,H(X)\bar\psi\rangle:= \int_{\rset^{3J}} \bar\psi(x_1,\ldots ,x_J)^*H(X)\bar\psi(x_1,\ldots ,x_J)dx_1\ldots dx_J
\]
is used here.
The mass of the nuclei, which are much greater than one (electron mass),  
are the diagonal elements in the diagonal matrix $M$.

The Ehrenfest dynamics (\ref{X_eq}-\ref{psi_eq}) is a Hamiltonian system
with the Hamiltonian
\[
M^{-1}\frac{|p|^2}{2} + \frac{1}{2}\langle \phi, H\phi\rangle =: H_E
\]
in the real variable $\big((X,\phi^r),( p, \phi^i)\big)=:(r_E,p_E)$, where $p=M\dot X$, 
\begin{equation}\label{real_part}
\phi=:\phi^r + i\phi^i \mbox{ is the decomposition into real $\phi^r$ and imaginary $\phi^i$ parts,}\end{equation}
$\langle \phi,\phi\rangle=2$ and 
$\bar\psi:=\phi/\langle\phi,\phi\rangle^{1/2}$; the normalization $\phi/\langle\phi,\phi\rangle^{1/2}$ is possible since the Schr\"odinger equation
\eqref{psi_eq} is linear and the norm $\langle \phi^t,\phi^t\rangle$ remains constant in time.
Consequently, the Ehrenfest dynamics conserves the energy $H_E$. 
The superscript $t$ denotes the time variable $X^t:=X(t)$, in the fast electron dynamics time scale.

The Ehrenfest dynamics can be derived from the time-independent Schr\"odinger equation for the full nuclei-electron system,
see Ref.   \cite{ASz}, and from the time-dependent Schr\"odinger equation, cf. Refs.   \cite{schutte,marx_hutter}.
The work Ref.   \cite{tully} describes  with examples some of the weaknesses and strengths of the Ehrenfest approximation.

The Ehrenfest dynamics can be coarse-grained by eliminating the 
electron dynamics and assuming the
electron wave function is in its ground state $\Psi_0$, satisfying the eigenvalue problem $H(X)\Psi_0(X)=\lambda_0(X)\Psi_0(X)$
for the lowest eigenvalue $\lambda_0(X)$, which reduces the Ehrenfest dynamics \eqref{X_eq}-\eqref{psi_eq} to
so called {\it Born-Oppenheimer dynamics} \cite{marx_hutter}
\[
\begin{split}
M\ddot X_{BO}^t &= -\partial_X\lambda_0(X_{BO}^t).
\end{split}
\]
The large mass $M\gg 1$ together with the bounded forces $\partial_X\lambda_0(X_{BO}^t)\sim 1$ 
make the dynamics slow and position increments are better studied
in the slower time scale, where $\tau:=M^{-1/2}t$ and $\hat\tau:= M^{1/2}\tau=t$,
\[
\frac{d^2}{d\tau^2} X_{BO}^{\hat\tau}= -\partial_X\lambda_0(X_{BO}^{\hat\tau}),
\]
since this dynamics does not depend on $M$; 
we will use greek letters $\tau,\sigma,\ldots$
to denote time in the slow scale and latin letters $t,s,\ldots$ for time in the fast scale.
The Born-Oppenheimer approximation 
leads to  %
accurate approximation of observables for the {\it time-independent
Schr\"odinger equation}, \begin{equation}\label{schrod_t_i}
\big(-(2M)^{-1}\Delta_X +H(X)\big)\Phi(X,\cdot)=E\, \Phi(X,\cdot),\end{equation} in
the case there is a spectral gap around the ground state:
\begin{equation}\label{observable_def}
|\int_{\rset^{3N}} g(X)\langle \Phi(X,\cdot),\Phi(X,\cdot)\rangle dX 
-\int_{\rset^{3N}} g(X)\rho_{BO}(dX)|=\mathcal O(M^{-1})
\end{equation}
for smooth functions $g$, see Ref.   \cite{ASz}; here  the energy
of the Born-Oppenheimer dynamics $\lambda_0(X_{BO}^t) + |\dot X_{BO}^t|^2/(2M)=E$ equals the Schr\"odinger eigenvalue. 
We use the big $\mathcal O$ and little $o$ notation for
\[\begin{split}
h(M) &=\mathcal O\big(f(M)\big) \iff \mbox{ $\exists\ C,M_0\in \rset$ such that } |h(M)|\le Cf(M) \mbox{ for } M> M_0,\\\
h(M) &=o\big(f(M)\big)\iff \lim_{M\rightarrow\infty} \frac{h(M)}{f(M)}=0.
\end{split}
\]
The first integral  in \eqref{observable_def} is the quantum mechanical measure of the nuclei-position 
{\it observable} $g(X)$
and $\int_{\rset^{3N}} g(X)\rho_{BO}(dX)$ is the the {\it micro canonical ensemble} average of $g(X_{BO})$,  
which would be equal to $\lim_{\mathcal T\rightarrow\infty} \mathcal T^{-1}\int_0^\mathcal T g(X_{BO}^{t}) dt$ if $X_{BO}$ would be ergodic. The two densities are normalized so that 
\[
1=\int_{\rset^{3N}} \langle \Phi(X,\cdot),\Phi(X,\cdot)\rangle dX =\int_{\rset^{3N}} \rho_{BO}(dX).
\]
The same result holds for the Ehrenfest dynamics replacing the Born-Oppenheimer dynamics;
however, this accuracy requires to know
the initial electron wave function  very precisely.  %
It seems reasonable to study randomness in the initial data of the electron wave function -- 
the modeling and accuracy with stochastic electron data  $\bar\psi$  in \eqref{psi_eq} is the purpose of this work.

The main inspiration for the stochastic model  here
is Refs.   \cite{ford_kac1}-  \cite{ford_kac2}, by Kac, Ford and Mazur, %
and in particular Ref.   \cite{zwanzig}, where Zwanzig  derives a Langevin equation for a heavy particle,
from a Hamiltonian system with the heavy particle 
coupled through a harmonic interaction potential to a heat bath particle system. 
The stochastics enters by having the heat bath degrees of freedom
initially  Gibbs distributed at a certain temperature, which after elimination of the heat bath degrees of freedom
yields a generalized Langevin equation for the heavy particle, including an integral operator for the friction term --
a so called memory term.
An assumption of a limiting continuous Debye
heat bath frequency distribution and certain linear weak coupling behavior 
then reduces the dynamics to a proper Langevin equation, where the integral
kernel becomes a point mass, i.e. an equation without memory terms.

 This work shows, in Section \ref{zwanzig}, that Zwanzig's model is closely related to the Ehrenfest Hamiltonian system
 and Sections \ref{bo_sektion}-\ref{sec_approx}  extends the ideas in Ref.   \cite{zwanzig} to
the {\it ab initio} Ehrenfest dynamics \eqref{X_eq} for nuclei
and the Schr\"odinger equation \eqref{psi_eq} for electrons (or other light particles).  
This means that the electron wave function plays the role of the "heat bath system",
the stochastics enters as Gibbs distributed initial data for the electron wave function $\bar\psi$,
and the Gibbs measure is parametrized by the temperature $T$.
The approximating {\it Langevin dynamics}, for the nuclei  positions $X_L^\tau$  in the slow time scale, takes the form
\[ %
\begin{array}{rl}
\dot X_L^\tau &=p_L^\tau\\
\dot p_L^\tau &=  -\partial_X\lambda_0(X_L^\tau)  - \hat Kp_L^\tau + (2T)^{1/2}  \hat K^{1/2} \dot W^\tau,\\
\end{array}
\] %
for the same temperature $T$,
where $W$ is the standard Wiener process in $\rset^{3N}$ and the positive $3N\times 3N$ friction/diffusion matrix 
$\hat K=\hat K(X_L^\tau)$ is a certain small function  (of the ground state $\Psi_0$) of order $M^{-1/2}$.

The purpose of this work is to study approximation  of Ehrenfest {\it observables} by Langevin observables, i.e. to estimate the difference 
\[
\mathbb E[g(X^{\hat\tau},p^{\hat\tau};X^0,p^0)]-\mathbb E[g(X_L^{\tau};X_L^0,p_L^0)], \mbox{ for bounded time $\tau$,}
\]  and 
\[
\lim_{\mathcal T\rightarrow\infty} {\mathcal T}^{-1}\int_0^{\mathcal T}\mathbb E[g(X^{\hat\tau},p^{\hat\tau};X^0,p^0)]-\mathbb E[g(X_L^{\tau};X_L^0,p_L^0)] \ d\tau
\]
for a given observable function $g$, where the expectation is with respect to the stochastic electron
initial data for the Ehrenfest dynamics $(X^t,p^t)$ and with respect to the Wiener process for the Langevin dynamics 
$(X^\tau_L,p_L^\tau)$; %
to find an accurate Langevin approximation includes 
to precisely determine the friction/diffusion matrix $\hat K$. %
The main idea is to use the Hamiltonian structure of the Ehrenfest dynamics to
formulate and determine stochastic molecular dynamics: there is a modeling part in Sections \ref{zwanzig} and \ref{Gibbs}, to formulate the stochastic initial data for the Ehrenfest dynamics, and an analysis part in Sections \ref{bo_sektion} and \ref{sec_approx}, to approximate the Ehrenfest dynamics by Langevin dynamics
in the canonical ensemble of constant number of particles
constant volume and constant temperature.
Theorem \ref{thm} determines a specific friction matrix so that Langevin dynamics
approximates
Ehrenfest dynamics for some observables including time-correlation,
in the canonical ensemble when the initial data for the electrons is a temperature
dependent stochastic perturbation of its ground state;
the accuracy is $o(M^{-1/2})$ on bounded time intervals (in the slow time scale of nuclei motion),
which makes the  $\mathcal{O}(M^{-1/2})$ small friction %
term visible.
Theorem \ref{long_time_thm} shows that the accuracy is $o(1)$ on unbounded time intervals,
assuming the time correlation length, of the first variation of the observable with respect to momentum,
 is at most of order $M^{1/2}$; this result, valid for long time approximation,
detects also the
$o(M^{-1/2})$ small diffusion term. 
The main assumption for the results is related to the reduction to a point mass for
the integral kernel in the friction term. In contrast to Ref.   \cite{zwanzig} no explicit expression of the
frequency distribution of the coupling is used. Instead the separation of slow nuclei dynamics
and fast electron dynamics, as $M\rightarrow\infty$, is used together with an assumption
that the electron eigenvalues form a continuum in the limit
as $M$ and $J$ tend to infinity, 
as described in \eqref{K_definition}. %
A second assumption is that the $\ell^1$ norm of the first variation of the observable is bounded, which
restricts the study to observables that are stable in the sense presented in Remark \ref{observable}.
The results also require the temperature to be low $T\ll 1$, all electron eigenvalues to be separated
and $ H$ to depend smoothly on $X$.

The particles with coordinates $x$, in \eqref{X_eq}-\eqref{psi_eq}, can also be interpreted as a heat bath of lighter particles consisting of both nuclei and electrons,
i.e. not necessarily only of electrons, 
so that the Langevin equation  also describes approximately the dynamics
of heavy so called Brownian particles. 
Theorems \ref{thm} and \ref{long_time_thm} therefore contribute to %
the central problem in statistical mechanics 
to show that  Hamiltonian dynamics of heavy particles, coupled to a heat bath of many lighter particles
with random initial data, can be approximately described by Langevin's equation, cf. Ref.   \cite{kadanoff}.
Such heat bath studies initiates from %
the pioneering work in Refs.   \cite{einstein},   \cite{langevin} 
and continues with more precise
heat bath models, based on harmonic interactions, in Refs.   \cite{ford_kac1,ford_kac2,zwanzig}.
More recently these  models of a heavy particle coupled to a heat bath are also used for numerical analysis
studies related to coarse-graining in molecular dynamics and  weak convergence analysis \cite{stuart1,stuart2,hald}, 
for strong convergence analysis \cite{gil_eric}, and for 
computational studies on nonlinear heat bath models \cite{stuart3,stuart4}.
Langevin's equation has  also been derived from a heavy particle colliding with
an ideal gas heat bath, where  the initial light particle positions are modeled by a Poisson point process
and  initial particle velocities are independent Maxwell distributed; 
the  heavy particle collides elastically with the ideal gas particles
and moves uniformly in between, see Refs.   \cite{lebo,lebo2}.

Five new issues here are:
\begin{itemize}
\item The slow nuclei dynamics compared to the fast electron
dynamics is exploited, 
to find a proper Langevin equation without using explicit heat bath frequencies.
\item The error analysis uses the residual in the
Kolmogorov equation, of the Langevin dynamics,
evaluated along the Ehrenfest dynamics, instead of the explicit solution available for harmonic oscillators.
\item A long time result uses exponential decay in time of the first variation of the observable with respect to
perturbations in the momentum.
\item %
An assumption of sufficiently low temperature makes certain quadratic forcing terms in the Ehrenfest momentum 
equation  negligible, to resemble the Zwanzig model. %
\item Why shall the electron initial data be
Gibbs distributed? The Ehrenfest dynamics with
the initial Gibbs distribution for the electrons is motivated from a stability 
and consistency argument in Section \ref{gibbs_der}.

\end{itemize}
The two central ideas in deriving Langevin dynamics from
coupling to a heat bath 
-- to find the friction mechanism in the heavy particle coupling to
the dynamics of the lighter particles and 
to find the diffusion from Gibbs fluctuations in initial data of the light particles -- were already the basis in 
Refs.   \cite{ford_kac1,ford_kac2} and   \cite{zwanzig}.

The outline of the paper is the following.
Section \ref{zwanzig} connects the Ehrenfest system to the Zwanzig model and 
Section \ref{Gibbs} presents a motivation for using a model with stochastic initial electron data based on the Gibbs distribution. 
Section \ref{bo_sektion} formulates the stochastic initial
data for the Ehrenfest model, including the normalization constraint, and states the two theorems
proved in Section \ref{sec_approx}.

\section{Zwanzig's Model and Derivation of Langevin Dynamics}\label{zwanzig}       %
This section reviews the heat bath
model of Zwanzig \cite{zwanzig}, with his derivation of stochastic Langevin dynamics from a Hamiltonian system,
related to the earlier work \cite{ford_kac1}.  The model consists of 
heavy particles interacting with many light particles which are initially
in a Gibbs distribution; in this sense, it models heavy particles in a heat bath of light particles.
A modification of the general formulation in Ref.
  \cite{zwanzig} is here to choose a special case closely resembling the
Ehrenfest dynamics. 
The model is as simple as possible to have the desired qualitative properties of
a system interacting with a heat bath.

The aim  is to show that the Ehrenfest model is 
closely related to the Zwanzig model and  to give some understanding
of simulating, at constant temperature, the coarse-grained molecular dynamics of the heavy particles
without resolving the lighter particles, using
Langevin dynamics.  It is an example how stochastics enter into a coarse-grained model
through elimination of some degrees of freedom in a determinstic model, described by
a Hamiltonian system.
The original model is time reversible while the coarse-grained
model is not. 

We consider Zwanzig's model \cite{zwanzig} with $N$ heavy particles %
and the particle positions  $X\in \rset^{3N}$  and momentum $p\in\rset^{3N}$ in a 
heat bath with $J$  light particles modes, with "position" coordinate %
$x\in\rset^{J}$ and dual "momentum" coordinate $q\in\rset^J$. %
Define, as a special case of  {\it Zwanzig's general model}, the Hamiltonian
\begin{equation}\label{ham_zwanzig}
H_Z(X,p,x,q):= \frac{1}{2} |p|^2 + \lambda(X)+ \frac{m}{2} \langle x-\hat\Psi(X),\hat H(x- \hat\Psi(X))\rangle + \frac{1}{2m}
\langle q,\hat H q\rangle
\end{equation}
where the operator $\hat H$ is a constant positive definite Hermitian $J\times J$ matrix, 
the coupling is represented by the function $\hat\Psi:\rset^{3N}\rightarrow\rset^J$, the bilinear form
$\langle\cdot,\cdot\rangle$ is now the Euclidian scalar product in $\rset^J$, the light particle mass is $m$, the heavy particle mass is one
and  $\lambda:\rset^{3N}\rightarrow\rset$ is a given potential. This Hamiltonian
yields the dynamics (in the slow time scale of the heavy particles)
\begin{eqnarray}
\ddot{X}_n^\tau &=& - \partial_{X_n}\lambda(X^\tau) +  \langle m\hat H(x-\hat\Psi(X^\tau)), \partial_{X_n}\hat\Psi(X^\tau)\rangle\quad n=1,\ldots, 3N
\label{z:heavy} \\ 
\dot x^\tau &=& m^{-1} \hat Hq^\tau\\
\dot q^\tau & =&- m\hat H (x^\tau-\hat\Psi(X^\tau)),
\label{z:linear}
\end{eqnarray}
where we use the notation $\Psi_0'(X):=\partial_X\Psi_0(X)$. 
To resemble the Ehrenfest dynamics, we define $y:=m^{-1}q$
and obtain
\[
\begin{split}
\dot x^\tau &= \hat H y^\tau\\
\dot y^\tau &= -\hat H(x^\tau-\hat\Psi(X^\tau))
\end{split}
\]
which using $\psi:=x-\hat\Psi(X) + i y$ shows the Zwanzig model written as a Schr\"odinger equation
coupled to heavy particle dynamics
\begin{equation}\label{zwanzig_eq}
\begin{split}
\ddot{X}^\tau &= - \lambda'(X^\tau) +  \Re \langle m\hat H\psi^\tau, \hat\Psi'(X^\tau)\rangle\\
i\dot \psi^\tau  &=  \hat H\psi^\tau  - i \hat\Psi'(X^\tau)\cdot \dot X^\tau,\\
\end{split}
\end{equation}
where $\Re w$ denotes the real part of $w\in \mathbb C$.

Let us compare the Zwanzig model to the Ehrenfest dynamics:
\begin{lemma}\label{ehrenfest_zwanzig_lemma}
The Ehrenfest dynamics \eqref{X_eq}-\eqref{psi_eq} can be written in the slow time scale as
\begin{equation}\label{lemma_ehren_eq}
\begin{split} %
\ddot X^\tau & = - \lambda'_0(X^\tau)   +2\Re\langle \tilde H\tilde\psi^\tau , \Psi_0'(X^\tau)\rangle  
-\langle\tilde\psi^\tau, \tilde H'(X^\tau)\tilde\psi^\tau\rangle,\\
i\dot{\tilde \psi}^\tau &= M^{1/2} \tilde H(X^\tau)\tilde\psi^\tau - i \Psi_0'(X^\tau)\cdot\dot X^\tau, \\ %
\end{split}
\end{equation}
using the definitions
\[
\begin{array}{lll}
&H\Psi_0  = \lambda_0\Psi_0, \quad &\mbox{eigenvalue}\\
&H_E(X,\phi_r; p,\phi_i):= \frac{|p|^2}{2} + \lambda_0(X)  
+ \frac{M^{1/2}}{2}\langle\phi, \tilde H\phi\rangle,  &\mbox{Hamiltonian}\\
&\tilde H := H - \lambda_0, \quad &\mbox{operator splitting}\\
&\psi := \frac{\phi}{\sqrt {\phi\cdot\phi}},\quad M^{1/2}/2=1/\langle\phi,\phi\rangle, &\mbox{normalization}\\
&\tilde\psi :=\psi- \Psi_0, &\mbox{wave function splitting,}\\
&\tau=M^{-1/2}t, & \mbox{slow time scale.}\\
\end{array}
\]
\end{lemma}
The lemma is proved in Section \ref{bo_sektion}, see \eqref{psi_eq_tilde} and \eqref{ehren_force}, by direct application of the definitions in the lemma.
We note that the Zwanzig dynamics \eqref{zwanzig_eq} and the Ehrenfest dynamics \eqref{lemma_ehren_eq}
 are similar
and by choosing 
\begin{equation}\label{ident}
\begin{split}
\hat\Psi(X) & = \Psi_0(X),\\
\hat H &= M^{1/2}\tilde H,\\
m &= 2M^{-1/2},\\
\lambda &= \lambda_0,
\end{split}
\end{equation}
in the case that $\tilde H$ is constant (in the subspace orthogonal to $\Psi_0$),
the two models are identical, since the 
quadratic interaction $-\langle\psi,\tilde H'\psi\rangle$ vanishes.
We also note that extending the Zwanzig model \eqref{ham_zwanzig} to a case when the matrix $\hat H:\rset^{3N}\rightarrow \rset^{J}\times\rset^J$
is a function of $X$ yields exactly  the Ehrenfest model \eqref{lemma_ehren_eq}, including the quadratic interaction.
We will see that the quadratic term is negligible for sufficiently low temperature. %
Consequently, the results for the general Zwanzig model are relevant for the Ehrenfest model;
this section reviews the results from Ref.   \cite{zwanzig}  and the other sections extends them 
to the Ehrenfest model (with non constant $\tilde H$).

\subsection{The initial data}\label{zwanzig_initial} We now study the Zwanzig model
\eqref{ham_zwanzig} and \eqref{zwanzig_eq}.
It seems reasonable to  assume that the many initial positions and velocities of the light particles are impossible to
measure and determine precisely. 
Clearly, to predict the dynamics
of the heavy particle some information of the light particle initial data
is necessary: we shall use an equilibrium probability distribution for the light
particles depending only on one parameter -- the temperature $T$ measured in units of the Boltzmann constant.
Section \ref{gibbs_der} presents  a motivation for stochastic sampling of $\psi(0)$
from the {\it Gibbs probability measure}
\begin{equation}\label{gibbs_zwan}
\begin{split}
&Z^{-1}\exp\big(-H_Z(X^0, p^0, x, q)/T\big)dx_1\ldots dx_J dq_1\ldots dq_J,\\
&Z:=\int_{\rset^{2J}} \exp\big(-H_Z(X^0, p^0, x, q)/T\big)dx_1\ldots dx_J dq_1\ldots dq_J.
\end{split}
\end{equation}
The distribution of the initial data $\psi(0)=x^0-\hat\Psi(X^0) + i q^0/m$, using for simplicity
$\hat\Psi(X^0)=0$, becomes clear in 
an orthonormal basis $\{\Psi_j\}_{j=1}^J$ of eigenvectors to $\hat H$, with the corresponding
eigenvalues $\{\lambda_j\}_{j=1}^J$, since writing  $\psi(0)=\sum_{j=1}^J\gamma_j\Psi_j$
implies 
\[
\frac{m}{2}\langle\psi^0,\hat H\psi^0\rangle
=\sum_{j=1}^J \frac{m\lambda_j}{2} |\gamma_j|^2= \sum_{j=1}^J \frac{m\lambda_j}{2}
(|\gamma_j^r|^2 +|\gamma^i_j|^2),\]
 so that
\begin{equation}\label{jacobi}
e^{-\frac{m}{2T}\langle \psi^0,\hat H\psi^0\rangle} dx_1\ldots dx_J dq_1\ldots dq_J
=e^{-\sum_{j=1}^J \frac{m\lambda_j}{2T} (|\gamma^j_r|^2 +|\gamma_i^j|^2)} d\gamma_1^r\ldots d\gamma_J^r 
d\gamma_1^i\ldots d\gamma_J^i,
\end{equation}
where it is used that the orthogonal transformation to the coordinates $x=\sum_j\gamma_j^r\Psi_j$
and $q=\sum_j\gamma_j^i\Psi_j$ has the Jacobian determinant equal to one
and $\gamma_j=\gamma_j^r + i\gamma_j^i$ is the decomposition into real and imaginary parts as in \eqref{real_part}.
We conclude that the complex valued $\gamma_j$ are all independent with independent normal  $N(0,2T/(m\lambda_j))$ distributed real and imaginary parts, $\gamma^j_r$ and $\gamma^j_i$.

\subsection{A generalized Langevin equation} Given the path $X$, the linear second equation in \eqref{zwanzig_eq} can by Duhamel's representation be solved explicitly, 
with the solution
\begin{equation}\label{zwanzig_pert}
\begin{split}
\psi(\tau) = - \int_0^\tau e^{-i(\tau-\sigma)\hat H} \hat\Psi'(X^\sigma) \dot X^\sigma d\sigma + \underbrace{e^{-i\tau\hat H} \psi(0)}_{z^\tau}.
\end{split}
\end{equation}
Insert this representation 
into the first equation of \eqref{zwanzig_eq} to obtain %
\begin{equation}\label{z:generalized}
\ddot{X}^\tau = - \lambda'(X^\tau)
- \int_0^\tau \langle m\hat H \cos\big((\tau-\sigma)\hat H\big) \hat\Psi'(X^\sigma) \dot X^\sigma, 
\hat\Psi'(X^\tau)\rangle d\sigma + 
\underbrace{\Re \langle m\hat H z^\tau, \hat\Psi'(X^\tau)\rangle}_{\zeta^\tau},
\end{equation}
where the last term has the components $\zeta_n=\Re \langle m\hat H z^\tau, \partial_{X_n}\hat\Psi(X^\tau)\rangle\in \rset^3, \ n=1,\ldots, N$.
To determine the covariance $\mathbb E_\gamma[\zeta^{\sigma}\otimes \zeta^{\tau*}]$, where $\mathbb E_\gamma$
denotes the expected value with respect to the initial data $\gamma$, note that the product of the fluctuation terms
\[
\big(\langle\ldots^\tau\rangle +\langle\ldots^\tau\rangle^*\big)
\big(\langle\ldots^\sigma\rangle +\langle\ldots^\sigma\rangle^*\big),
\]
where $\langle\ldots^\tau\rangle := \langle m\hat H z^\tau/2,\Psi'(X^\tau)\rangle$, yields four terms.
This sum of four terms can be written as follows;
use first that the initial data is $\psi^0=\sum_j\gamma_j\Psi_j$,
where $\{\gamma_j\}_{j=1}^J$ forms
a set of independent normal distributed random variables,
and then that $\{\Psi_j\}_{j=1}^J$ is an orthonormal basis to obtain
\begin{equation}\label{zwanzig_kovar}
\begin{split}
\mathbb E_\gamma[\zeta^{\sigma}\otimes \zeta^{\tau*}]&=
\mathbb E_\gamma \big[\frac{1}{2}\Re \langle m\hat H e^{-i\sigma\hat H}\psi^0,\hat \Psi'(X^\sigma)\rangle \langle \hat \Psi'(X^\tau), m\hat H e^{-i\tau\hat H}\psi^0\rangle\big]\\
&\qquad +
\mathbb E_\gamma \big[\frac{1}{2}\Re \langle m\hat H e^{-i\sigma\hat H}\psi^0,\hat \Psi'(X^\sigma)\rangle 
\langle m\hat H e^{-i\tau\hat H}\psi^0,\hat \Psi'(X^\tau) \rangle\big]\\
&\simeq\frac{1}{2}\Re \sum_{j,k}\langle m\lambda_j \Psi_j, e^{i\sigma\hat H}\hat \Psi'(X^\sigma)\rangle 
\langle  m\hat H e^{i\tau\hat H}\hat \Psi'(X^\tau), \Psi_k\rangle 
\underbrace{\mathbb E_\gamma[\gamma_j^*\gamma_k]}_{=4T\delta_{jk}/(m\lambda_j)}\\
&\qquad +
\frac{1}{2}\Re \sum_{j,k}\langle m\lambda_j \Psi_j, e^{i\sigma\hat H}\hat \Psi'(X^\sigma)\rangle 
\langle   \Psi_k,m\hat H e^{i\tau\hat H}\hat \Psi'(X^\tau),\rangle 
\underbrace{\mathbb E_\gamma[\gamma_j^*\gamma_k^*]}_{=0}\\
&= 2T\Re\sum_j \langle \Psi_j,  e^{i\sigma\hat H}\hat \Psi'(X^\sigma)\rangle 
\langle  m\hat H e^{i\tau\hat H}\hat \Psi'(X^\tau), \Psi_j\rangle\\
&=2T\langle m\hat H \cos \big((\tau-\sigma)\hat H\big)\Psi'(X^\sigma),\Psi'(X^\tau)\rangle;
\end{split}
\end{equation}
we have in the second step, where $a\simeq b\iff a=b(1+o(1))$,  used
that $X$ depends on $\gamma_n$ in such a way that the coupling between $X$ and $\gamma_n$
leads to lower order terms $\mathcal O(M^{-1/2})$, which is verified in Lemma \ref{main_lemma}.
We see that 
the {\it covariance} of the Gaussian process, $\zeta:[0,\infty)\times\{\mbox{probability outcomes}\}\rightarrow \rset^{3N}$,
\begin{equation}\label{zwanzig_friktion}
\begin{split}
\mathbb E_\gamma[(\zeta^\sigma \otimes \zeta^{\tau*})_{kl}]\simeq 2T \langle m\hat H \cos\big((\tau-\sigma)\hat H\big) \partial_{X_k}\hat\Psi(X^\sigma),\partial_{X_l}\hat\Psi(X^\tau)\rangle =:
2T f_{kl}(\tau-\sigma),\\
\end{split}
\end{equation}
is the multiple $2T$ of the integral kernel for the friction term in the generalized Langevin equation \eqref{z:generalized}, forming
a version of {\it Einstein's fluctuation-dissipation} result.

\subsection{A pure Langevin equation}
The next step in Zwanzig's modeling is to derive a pure Langevin equation for a special choice of
oscillation frequencies $\hat H$ and coupling $\hat\Psi$. Write 
$\hat\Psi'(X) =:\sum_{j=1}^J \hat\Psi'_j(X)\Psi_j$  to obtain
\[
\begin{split}
& \langle \hat H \cos\big((\tau-\sigma)\hat H\big) \partial_{X_k}\hat\Psi(X^\sigma),\partial_{X_l}\hat\Psi(X^\tau)\rangle\\
& =\sum_{j=1}^J\lambda_j\cos\big((\tau-\sigma)\lambda_j\big) \partial_{X_k}\hat\Psi_j(X^\sigma)^*\partial_{X_l}\hat\Psi_j(X^\tau).
\end{split}
 \]
 Zwanzig's derivation is for $X\in\rset$. Here we extend his setting to $X\in\rset^{3N}$ which will give a
 friction matrix and not only a scalar friction coefficient.
 There are two steps to obtain an integral kernel that reduces to a point mass.
 First, as $J\rightarrow \infty$, the distribution of the eigenvalues assumes a continuum distribution of frequencies. 
 Then the coupling function $\hat\Psi'$ is chosen so that the integral over all frequencies, in the limit
 as a cut-off frequency tends to infinity, is the Fourier transform of a constant, which is the point mass.
Zwanzig assumes that the frequencies $\lambda_j$ %
are distributed so that the sum
over the light particle modes  tends to an integral over frequencies with a {\it Debye distribution}, i.e. for any continuous function $h$
\begin{equation}\label{debye_fordel}
J^{-1}\sum_{j=1}^J h(\lambda_j) \rightarrow \int_0^{\lambda_d} h(\lambda) \frac{3\lambda^2}{\lambda_d^3} d\lambda.
\end{equation}
Let the coupling between the heavy particle and the heat bath be linear and assume there is a constant 
$\kappa$ so that there holds
\[
\partial_{X_k}\hat\Psi_{j}(X^\sigma) =\kappa^{1/2}(\frac{2}{Jm\pi})^{1/2} (\frac{\lambda_d}{\lambda_j})^{3/2}e^{-i\alpha_k\lambda_j} %
\]
where we assume that the parameters $\alpha_k\in\rset $ for $k=1,\ldots ,3N$ satisfy
\begin{equation}\label{alfa_cond}
|\alpha_k-\alpha_l|\rightarrow \infty \mbox{  as } \lambda_d\rightarrow\infty \mbox{ for $k\ne l$}.
\end{equation}
Then we have
\begin{equation}\label{f_redukt}
\begin{split}
f_{kl}(\tau) &\rightarrow \frac{2\kappa}{\pi} \int_0^{\lambda_d} \cos(\tau \lambda) e^{i(\alpha_k-\alpha_l)\lambda} d\lambda\\
&=\frac{\kappa}{2} \frac{2}{\pi}\Big(\frac{\sin \big(\lambda_d (\tau+\alpha_k-\alpha_l)\big)}{\tau+\alpha_k-\alpha_l} + \frac{\sin \big(\lambda_d (\tau-\alpha_k+\alpha_l)\big)}{\tau-\alpha_k+\alpha_l}\Big),\quad \mbox{ as } J\rightarrow\infty.
\end{split}\end{equation}
The phase shift $e^{-i\alpha_k\lambda_j}$ is introduced to have asymptotically vanishing correlation for $k\ne l$.
The integral kernel $f(\tau)$ converges weakly to $\kappa$ times the identity matrix as $\lambda_d\rightarrow\infty$,
since the change of variables $\omega=\lambda_d \tau$ yields
 \begin{equation}\label{delta}
\begin{split}
&\int_0^\infty \frac{\sin (\lambda_d \tau)}{\tau} h(\tau) \, d\tau
=\int_0^\infty \frac{\sin \omega}{\omega} h(\frac{\omega}{\lambda_d})d\omega
\rightarrow \underbrace{\int_0^\infty \frac{\sin \omega}{\omega}\, d\omega}_{\pi/2}\  h(0),\\
&\frac{\sin \big(\lambda_d (\tau+\alpha_k-\alpha_l)\big)}{\tau+\alpha_k-\alpha_l}\rightarrow 0\ , \mbox{ for } k\ne l,
\mbox{ by \eqref{alfa_cond}},
\end{split}
\end{equation}
which formally leads to the Langevin equation
\begin{equation}\label{langevinen}
\begin{split}
dX^\tau&=p^\tau \, d\tau,\\
dp^\tau &= \big(- \lambda'(X^\tau) - \hat K p^\tau\big)d\tau  + \sqrt{2T\hat K}\,  dW^\tau,
\end{split}
\end{equation}
as $J,\lambda_d\rightarrow \infty$ %
where $W$ is the standard Wiener process with independent components in $\rset^{3N}$
and $\hat K=\kappa I$ is  $\kappa$ times the $3N\times 3N$ identity matrix;
Theorems \ref{thm} and \ref{long_time_thm} verify a related limit carefully for the Ehrenfest model by giving precise conditions, without using the Debye distribution and the linear coupling.

This Langevin equation satisfies the ergodic limit
\[
\lim_{\mathcal T\rightarrow\infty} \mathcal T^{-1}\int_0^\mathcal T g(X^t,p^t) \, dt = 
\int_{\rset^{6N}} g(X,p) \mu(dX,dp)
\]
for instance if the invariant measure
\[
 \mu( dX, dp)=\frac{e^{-(|p|^2/2 + \lambda(X))/ T}\,  dX\, d p}{\int_{\rset^{6N}} e^{-(|p|^2/2 + \lambda(X))/ T}\,  dX\, d p}\, , 
 \]
 exists, $\lambda$ and $\hat K$ are smooth and $\hat K$ has full rank,  cf. Ref.   \cite{tony_book}.  
 The probability measure  $\mu$ is then also the unique solution of the corresponding Kolmogorov forward equation
 and it is the heavy particle marginal distribution of the Gibbs distribution 
 \[
 \frac{e^{-H_Z(X,p,x,q)/ T}\, dX\, dp\, dx\, dq}{\int_{\rset^{6 N+2J}}
 e^{- H_Z(X, p,x,q)/T}\,  dX\, dp\,  dx\, dq}
 \]
in \eqref{gibbs_zwan}.
We conclude that sampling the light particles from the Gibbs distribution, conditioned on the heavy particle coordinates,
 formally leads  time asymptotically to having the heavy particle in the heavy particle marginal of the Gibbs distribution:
 this fundamental stability and consistency property is in some sense unique to the Gibbs distribution, as explained in the
 next section.
 
 The purpose of this work is to apply Zwanzig's derivation to the Ehrenfest model of
 nuclei and electrons, acting as heavy and light particles, respectively.
 As we have seen the two models are similar and the difference is that
 the operator $\tilde H$ is a function of $X$ in the Ehrenfest model.
 To still be in a setting close to the Zwanzig model, we assume
 therefore that the temperature is low, so that the ground state dominates over small fluctuations, 
namely $\sum_{j=1}^J T/(\lambda_j(X^0)-\lambda_0(X^0))\rightarrow 0$ as $M\rightarrow \infty$ as described in \eqref{T_vilkor}. %
Three new issues in the Ehrenfest model %
are the lack of an explicit solution for the electron wave function,
the non linear coupling of the wave function to the nuclei
and the additional constraint to have a normalized wave function $\langle \psi^0,\psi^0\rangle=1$,
changing the initial distribution.
The extension to a non constant operator $\tilde H$ requires different mathematical tools
as compared to the case with an explicit solution. Here we will use estimates of the
residual  in the Kolmogorov backward 
equation for the Langevin equation evaluated along the Ehrenfest dynamics. A
main result is to determine
the diffusion/friction matrix without assuming explicit  properties of the spectrum of $\tilde H$,
but instead use the separation of time scales in the nuclei and electron dynamics
and %
a continuous spectrum assumption of the limit $\tilde H_\infty$ of $\tilde H$ as $M,J\rightarrow\infty$.

To explain how this works, the idea is first sketched in the simpler setting
of a constant $\tilde H$, i.e. the Zwanzig model using the identification \eqref{ident}.
The friction term, from the generalized Langevin equation \eqref{z:generalized}, is  assumed to satisfy
the equalities 
\begin{equation}\label{k_z}
\begin{split}%
&\lim_{M\rightarrow\infty} %
2M^{1/2}\int_0^\tau \Re\langle  \cos (\sigma M^{1/2}\tilde H) 
\frac{d}{d\tau}\Psi_0(X^{\tau-\sigma }),\tilde H\partial_X\Psi_0(X^\tau)\rangle d\sigma \\ %
&=%
\lim_{M\rightarrow\infty} 2\int_0^{M^{1/2}\tau} %
\Re\langle \cos (\hat \sigma \tilde H) 
\frac{d}{d\tau}\Psi_0(X^{ \tau-M^{-1/2}\hat \sigma }), 
 \tilde H\partial_X\Psi_0(X^\tau)\rangle %
 d\hat \sigma \\
& =2\int_0^\infty \lim_{M\rightarrow\infty} %
\Re\langle 
\underbrace{1_{\hat \sigma \le M^{1/2}\tau}
\tilde H\cos (\hat \sigma \tilde H) 
\frac{d}{d\tau}\Psi_0(X^{\tau-M^{-1/2}\hat \sigma })}_{=:\, \Gamma_M(\hat\sigma)}, 
\partial_X\Psi_0(X^\tau)\rangle   \,  %
d\hat \sigma \\
& =2\int_0^\infty %
\langle \tilde H_\infty\cos({\hat \sigma \tilde H_\infty}) \partial_X\Psi_0(X^{\tau}), 
\partial_X\Psi_0(X^\tau)\rangle  %
d\hat \sigma \, \dot X^\tau\\
&=: K(X^\tau)\dot X^\tau
\end{split} %
\end{equation}
where the first equality applies the change of variables $\hat\sigma=M^{1/2}\sigma$,
the second follows by dominated convergence, if the sequence $\Gamma_M$ converges pointwise 
and is bounded by an
$L^1(\rset)$-function $|\Gamma_M(\hat\sigma)|\le \Gamma(\hat\sigma)$, and the third equality uses 
the pointwise convergence of $\tilde H\cos (\hat \sigma \tilde H) 
\frac{d}{d\tau}\Psi_0(X^{\tau-M^{-1/2}\hat \sigma })$, based on 
$\sigma\mapsto \frac{d}{d\tau}\Psi_0(X^{\tau-\sigma})$
being continuous, $X^{\tau-M^{-1/2}\hat\sigma}\rightarrow X^\tau$, and an assumption on 
the spectrum of $\tilde H\rightarrow\tilde H_\infty$ as $J\rightarrow\infty$, for example the Debye distribution \eqref{debye_fordel}.
In \eqref{fourier_representation} the last integral is  described by the spectrum. 
Note that $K$ is a $3N\times 3N$ matrix and that
the friction term in the Langevin equation \eqref{langevinen}, approximating Ehrenfest dynamics,
becomes small $\hat K=M^{-1/2}K=\mathcal O(M^{-1/2})$ due to the definition of
$\hat K$, in \eqref{z:generalized}-\eqref{zwanzig_friktion} and \eqref{f_redukt}, without the factor $M^{1/2}$ in the left hand side of \eqref{k_z}.

\section{The Gibbs Distribution motivated from Dynamic Stability}\label{gibbs_der}\label{Gibbs}

At the heart of Statistical Mechanics is the Gibbs distribution
\[
\frac{e^{-H(Y,Q)/T} dY dQ}{\int_{\rset^{6N}} e^{-H(Y,Q)/T} dYdQ}
\]
for an equilibrium probability distribution of a Hamiltonian dynamical system
\begin{equation}\label{ham_yq}
\begin{split}
\dot Y^\tau  &= \partial_Q H(Y^\tau ,Q^\tau )\\
\dot Q^\tau  &= -\partial_Y H(Y^\tau ,Q^\tau )\\
\end{split}
\end{equation}
in the canonical ensemble of constant number of particles $N$, volume and temperature $T$.
Every book on Statistical Mechanics gives a motivation of the Gibbs distribution,
often based on entropy considerations, cf. Refs.   \cite{feynman},   \cite{lanford}. %
The purpose of this section is to present a different motivation of
the Gibbs distribution, based on dynamic stability and consistency, for large systems with many heavy particles
where the energy can be split into a sum of a dominating heavy (nuclei) term and smaller light (electron) term.

Consider a Hamiltonian system with
heavy and light particles, with position $Y=(X,x)$, momentum $Q=(p,q)$ and a Hamiltonian 
$H=H_h(X,p) + H_l(X,x,q)$; an example is  \eqref{ham_zwanzig} where
\begin{equation}\label{heavy_light}
H_h=|p|^2/2 + \lambda(X) \mbox{ and } H_l=m\big\langle x-\hat\Psi(X),\hat H\big(x-\hat\Psi(X)\big)\big\rangle/2
+\langle q,\hat Hq\rangle/(2m).
\end{equation}
Assume that it is impractical or impossible to measure and determine the initial data $(x^0,q^0)$ for the light particles.
Clearly it is necessary to give some information on the data to determine the solution at a later time.
In the case of molecular dynamics it is often sufficient to know the distribution of the particles
to determine thermodynamic relevant properties, as e.g. the pressure-law.
 We saw in Section \ref{zwanzig} that if the light particles have an initial probability distribution corresponding
the  Gibbs distribution conditioned on the heavy particle, then the invariant distribution for the heavy particle is unique 
(in the limit of the Langevin equation) and given by the Gibbs distribution
for the heavy particle 
\[
\frac{e^{-H_h(X,p)/T} dX dp}{\int_{\rset^{6N}} e^{-H_h(X,p)/T} dX dp};
\]
this invariant Gibbs distribution is for constant $\hat H$ equal to the
the heavy particle {\it marginal distribution}
\[
 \frac{\int_{\rset^{2J}}e^{-H/T} dX dpdxdq}{\int_{\rset^{6N+2J}} e^{-H/T} dX dpdxdq}
\]
of the Gibbs measure for the full Hamiltonian integrated over the light particle phase-space.
This stability that an equilibrium distribution of light particles leads
to the marginal distribution of the heavy particles holds only  for the Gibbs distribution in the sense 
 we shall verify below, for systems where the heavy particle energy dominates
 as the number of heavy particles tends to infinity.
 Define the set $\mathcal S$ of equilibrium measures that have this desired stability and consistency property
 more precisely as follows:
\begin{itemize} 
\item[(S)]
an equilibrium measure $\nu(X,p,x,q)dXdpdxdq$ belongs to $\mathcal S$ if
 the dynamics of the heavy particles, with the light particles initially distributed
 according to $\nu(X^0,p^0,\cdot,\cdot) dxdq$ (the equilibrium distribution conditioned on the heavy particle initial data),
 asymptotically after long time ends up with the 
 heavy particles distributed according to $\int_{\rset^{2J}}\nu(X,p,x,q)dxdq \, dXdp$ 
 (the heavy particle marginal of the original equilibrium measure).
 \end{itemize}
Consequently the heavy particle behavior over long time, using $\nu\in\mathcal S$,
 is consistent with the assumption to start the light particles with 
 the equilibrium distribution $\nu$. Below we motivate why $\mathcal S$ only contains the Gibbs measure
 under certain assumptions.
 It is in fact the uniqueness of the Gibbs initial probability distribution
 that makes a stochastic model of the dynamics useful: if we would have to seek the initial distribution among
 a family of many distributions we could not predict the dynamics in a reasonable way.

The consistency property (S) is in an abstract
setting related to the study on Gibbs measures for
lattice systems in Ref.   \cite{lanford}, 
where dynamics and conditioning with respect to initial heavy  particles is 
replaced by invariance for
conditioning with respect to complement sets on the lattice, as the lattice growths to infinity. 
 
 To study this uniqueness of the Gibbs density, we consider first all equilibrium densities of 
 the Hamiltonian dynamics and then use the consistency check (S)
 to rule out all except the Gibbs density.
 There are many equilibrium distributions for a Hamiltonian system: 
 the {\it Liouville equation} (i.e. the Fokker-Planck equation for zero diffusion)
 \[
\underbrace{\partial_t f(H)}_{=0} + \partial_Y(\partial_Q H f(H)) -\partial_Q(\partial_Y H f(H))=0
\]
shows that any positive (normalizable) function
 $f$, depending only on the Hamiltonian $H$ and not on time, 
is an invariant probability distribution
 \[
dG:=\frac{f\big(H(Y,Q)\big) dY dQ}{\int_{\rset^{6N}} f\big(H(Y,Q)\big)dYdQ}
\]
for the Hamiltonian system \eqref{ham_yq}. There may be other invariant solutions
which are not functions of the Hamiltonian but these are not considered here.
Our basic question is now --
which of these functions $f$  have the fundamental property
that  their light particle  distribution generates a unique invariant measure
given by the heavy particle marginal distribution?
We have seen that the Gibbs distribution  is such a solution for the Zwanzig model. Are there other?

Write $H= H_h+ H_l$
and assume that the heavy particle Hamiltonian $H_h$
dominates the light particle Hamiltonian $H_l$, so that
\[ %
\frac{H_l}{H_h} \rightarrow 0 \mbox{ in $L^1(dG)$ as the number of heavy particles tend to infinity.}
\] %
In the Zwanzig setting with $N$ heavy particles and  $J$ light particles modes
the light particle energy and the equilibrium heavy kinetic energy
are proportional to the temperature, $\mathbb{E}[H_l]=TJ$ and $\mathbb{E}[|p|^2/2]=3TN$, 
while the potential energy $\lambda(X)$ is typically an extensive \cite{kadanoff} variable
 proportional to $N$, independent of $T$.
If the temperature is sufficiently low, we would then need that $TJ/N\rightarrow 0$ as $N\rightarrow \infty$
for the ratio of light and heavy particle expected energy to vanish asymptotically.
In Section \ref{stok_init_sec} we use a sum of $J$ bounded random variables as initial data for the Ehrenfest electron dynamics
to obtain also $\|H_l\|_{L^\infty}=\mathcal O(TJ)$ and $\|H_l/H_h\|_{L^\infty}=\mathcal O(TJ/N)$.
Therefore we may use the more restrictive assumption 
\begin{equation}\label{H_bound}
\|\frac{H_l}{H_h}\|_{L^\infty} \rightarrow 0 \mbox{ as $N\rightarrow\infty$.}
\end{equation}

Let
\[
-\log f(H)=: g(H)
\]
and consider perturbations of the Gibbs distribution in
the sense that the function $g$ satisfies  for a constant $C$
\begin{equation}\label{g_bound}
\begin{split}
\overline{\lim}_{H\rightarrow \infty} \Big|\frac{g''(H)H}{g'(H)}\Big|\le C\\
\overline{\lim}_{H\rightarrow \infty} \Big|\frac{g'(H)H}{g(H)}\Big|\le C
\end{split}
\end{equation}
for instance, any monomial $g$ satisfies  \eqref{g_bound}.
Taylor expansion yields for some $\alpha\in (0,1)$
\[
\begin{split}
-\log f(H)&= g(H_h+H_l) \\
&=
g(H_h) + H_l\big(g'(H_h) +2^{-1}g''(H_h + \alpha H_l) H_l\big)\\
\end{split}
\]
and \eqref{H_bound} and \eqref{g_bound} imply the leading order term
\begin{equation}\label{H_l-expansion}
-\log f(H) \simeq g(H_h) + H_lg'(H_h),
\end{equation}
as explained in Remark \ref{H_asympt};
in words, this means that the heavy particle energy dominates and
acts as a heat bath to find the distribution for 
the light particles.
Define the constant 
\begin{equation}\label{T-def}
T=1/g'\big(H_h(X_0,p_0)\big).
\end{equation}
The light particle distribution is then initially approximately given by
\begin{equation*} %
\frac{e^{-H_l/T} dx dq}{ \int e^{-H_l/T} dx dq}\, .
\end{equation*}
Neglecting the higher order terms, this leading order initial Gibbs distribution, with the temperature $T=1/g'\big(H_h(X_0,p_0)\big)$, applied to the Zwanzig model (or the Ehrenfest model) leads time asymptotically,
by the derivation of \eqref{langevinen} (alternatively Theorem \ref{long_time_thm}),  to the 
heavy particle equilibrium distribution 
\begin{equation}\label{h_equil}
\frac{e^{-H_h/T} dX dp}{\int e^{-H_h/T} dX dp}\, .\end{equation}
On the other hand, the equilibrium density $f$ has by \eqref{H_bound} and \eqref{g_bound} the  expansion
\[
\begin{split}
-\log f(H)&= g(H_h+H_l) \\
&=
g(H_h) + g'(H_h + \alpha H_l) H_l\\
&\simeq g(H_h),
\end{split}
\] 
which asymptotically as $N\rightarrow\infty$ (obtained as in \eqref{H_l-expansion}) becomes to the heavy particle marginal distribution
\begin{equation}\label{h_marg}
\frac{e^{-g(H_h)} dX dp}{\int e^{-g(H_h)} dX dp}.\end{equation}
The consistency requirement to have the heavy particle distribution \eqref{h_equil}
equal to the heavy particle marginal distribution \eqref{h_marg}
implies that
\[
g(H_h)= H_h/T. %
\]
We conclude that the quotient $-H/\log f(H)$ is constant, where $-H/\log f(H)=T$ is called the temperature,
 in units of the Boltzmann constant,
and we have established our motivation of the Gibbs density $f(H)=e^{-H/T}$.

A more complete derivation that would prove the validity of the Gibbs distribution along this line would need
to study the Ehrenfest or Zwanzig dynamics under the 
perturbation caused by having the initial light particle distribution in the distribution $dG$.
The presentation here is restricted to the simpler (and more vague) motivation based on using the obtained leading order (Gibbs) distribution of the initial light particles. However the weak convergence analysis in 
Theorems \ref{thm} and \ref{long_time_thm} is possible to generalize to perturbed initial Gibbs distributions having the right asymptotic covariance; the critical issue is
to modify \eqref{u_koppling}-\eqref{fluct_diss}, which is possible since it already is based on the leading order distribution. 

\begin{remark}\label{H_asympt}
We have for $\|H_l/H_h\|_{L^\infty}$ sufficiently small and $H_h$ sufficiently large
\begin{equation}\label{A_lim}
A:=\sup_{\alpha\in (0,1), H_l<cH_h}  |\frac{g''(H_h+\alpha H_l)(H_h+\alpha H_l)}{g'(H_h)}|\le \frac{2C}{1-2CH_l/H_h}
\end{equation}
since \eqref{g_bound} implies 
\[
\begin{split}
A &=\sup_{\alpha\in (0,1), H_l<cH_h} %
|\frac{g''(H_h+\alpha H_l)(H_h+\alpha H_l)}{g'(H_h+\alpha H_l)}|
|\frac{g'(H_h+\alpha H_l)}{g'(H_h)}|\\
&\le 2C(1+|\frac{g''(H_h+\alpha_1 \alpha H_l) \alpha H_l}{g'(H_h)}|)\\
&\le 2C(1+A\frac{H_l}{H_h})
\end{split}
\]
for some $\alpha_1\in (0,1)$,
so that 
\[
A \le \frac{2C}{1-2CH_l/H_h} . %
\]
Therefore \eqref{A_lim} yields the expansion
\[
\begin{split}
f &= e^{g(H_h) + H_l\big(g'(H_h) +2^{-1}g''(H_h + \alpha H_l) H_l\big)}\\
&= e^{g(H_h)}  e^{H_l g'(H_h)(1+\mathcal O(H_l/H_h))}\\
&= e^{g(H_h)}  e^{H_l g'(H_h)} 
+ e^{g(H_h)}  e^{H_l g'(H_h)}\mathcal O(H_l/H_h)\\
&= e^{g(H_h)}  e^{H_l g'(H_h)} 
+ f\mathcal O(H_l/H_h)\\
\end{split}
\]
with the leading order term $e^{g(H_h)}  e^{H_l g'(H_h)} $, provided $H_lg'(H_h)\gtrsim 1$
and $|\ H_l/H_h\|_{L^\infty} \rightarrow 0$ as $N\rightarrow\infty$ by assumption \eqref{H_bound}.
\end{remark}

\section{Data for the Ehrenfest Model and the Main Results}\label{bo_sektion}
At low temperature one expects that the fast electron dynamics, compared to the
slower nuclei in the Ehrenfest dynamics, 
yields an electron wave solution that is almost in its {\it ground state} $\Psi_0$, which solves the electron eigenvalue problem 
\begin{equation}\label{ground_state}
H\Psi_0=\lambda_0\Psi_0,
\end{equation}
and is normalized $\langle \Psi_0,\Psi_0\rangle=1$;
here $\lambda_0=\lambda_0(X^t)$ is the smallest
eigenvalue of $H=H(X^t)$ in the considered Hilbert space (a subset of $L^2(dx)$). The function  \[
\hat\Psi^t:=e^{-i\int_0^t\lambda_0(X^s)ds}\Psi_0(x;X^t)\]
satisfies
\[
i\,\dot{\hat \Psi}^t - H\hat\Psi^t=i\, e^{-i\int_0^t\lambda_0(X^s)ds}\frac{d}{dt}\Psi_0(x;X^t),
\]
so that if the nuclei do not move, the wave function $\hat\Psi$ solves the  time-dependent Schr\"odinger equation;
and if they move slowly, i.e. the $L^2$-norm $\|\dot\Psi_0\|$ is small, the function $\hat\Psi$ is an
approximate solution to the Ehrenfest dynamics. The approximation $(X^t,\hat\Psi)$, 
with $\hat\Psi$ replacing $\bar\psi$ in \eqref{X_eq}, is called {Born-Oppenheimer
molecular dynamics} \cite{marx_hutter,hagedorn} and it approximates observables of
the time-independent Schr\"odinger equation for the electron-nuclei system
with accuracy $\mathcal{O}(M^{-1})$, when there is a spectral gap around the electron ground state, see Ref.   \cite{ASz}.
Here we study  molecular dynamics when the electron states $\bar\psi$ are randomly perturbed from the ground state,
with a Gibbs distribution at positive  temperature. To follow the perturbation of the
ground state, make the transformation
\begin{equation}\label{psi_bar_ekv}
\bar\psi(t,x)=e^{-i\int_0^t\lambda_0(X^s)ds}\psi(t,x)
\end{equation}
which implies that $\psi$ solves the Schr\"odinger equation
\begin{equation}\label{psi_tilde_eq}
i\frac{d}{dt} \psi^t = \underbrace{\big(H(X^t)- \lambda_0(X^t)\big)}_{=:\tilde H(X^t)}\psi^t,
\end{equation}
with the translated Hamiltonian $\tilde H$,
and then use the Ansatz
\begin{equation}\label{psi_tilde}
\psi= \tilde\gamma_0\Psi_0+\tilde\psi
\end{equation}
for some constant $\tilde\gamma_0\simeq 1$
and expect $\tilde \psi:[0,\infty)\times \rset^{3J}\rightarrow\mathbb{C}$ to be small.
The Ansatz implies that $\tilde\psi$ solves
\begin{equation}\label{psi_eq_tilde}
i \frac{d}{dt} \tilde\psi={\tilde H(X^t)}\tilde\psi-i\tilde\gamma_0\frac{d}{dt}\Psi_0,
\end{equation}
which yields the solution representation (of the same qualitative form as \eqref{zwanzig_pert} in Zwanzig's model)
\[
\tilde\psi^t= \tilde S_{t,0}\tilde\psi^0
-\tilde\gamma_0\int_0^t  \tilde S_{t,s}\underbrace{
\dot {\Psi}_0^s}_{\partial_X\Psi_0\dot X} ds,
\]
with the solution operator $\tilde S$ defined by
\begin{equation}\label{solution_operator}
\tilde S_{t,s}\varphi^s:= \varphi^t\end{equation}
for the solution in the fast time scale
\[
i\frac{d}{dt}\varphi^t=\tilde H(X^t)\varphi^t \quad t>s.
\]
The first term in the representation depends only on the initial data and the second term depends only
on the residual $\dot\Psi_0$. This splitting, inserted into the equation \eqref{X_eq} for the nuclei,
eliminates formally the electrons and generates fluctuations, from stochastic initial data $\tilde\Psi^0$,
and friction, through the coupling to $\dot\Psi_0=\partial_X\Psi_0\dot X$, as explained in Section \ref{zwanzig}. 

\subsection{The Force on Nuclei}\label{sec:force}
It is convenient to split the force on nuclei in \eqref{X_eq} into two parts, using the
definition of $\psi$ in \eqref{psi_bar_ekv} and the normalization $\langle\psi,\psi\rangle=1$,
\[
\begin{array}{rl}
\langle \bar\psi,\partial_X H(X)\bar\psi\rangle&=
\langle \psi,\partial_X H(X)\psi\rangle\\
&= \langle \psi,\partial_X\lambda_0(X)\psi\rangle
+ \langle \psi,\partial_X \big(H(X)-\lambda_0(X)\big)\psi\rangle\\
&=\partial_X\lambda_0(X)
+ \langle \psi,\partial_X \tilde H(X)\psi\rangle.
\end{array}
\]
With the Ansatz $\psi=\tilde\gamma_0\Psi_0+\tilde\psi$,  for a constant $\tilde\gamma_0\simeq 1$,
the second term in the
nuclear force becomes
\[
\begin{array}{rl}
\langle \psi,\partial_X \tilde H(X)\psi\rangle &= |\tilde\gamma_0|^2 \langle \Psi_0,\partial_X \tilde H(X)\Psi_0\rangle\\
&\qquad +\langle \tilde\psi,\partial_X \tilde H(X)\tilde\gamma_0\Psi_0\rangle
+\langle \tilde\gamma_0\Psi_0,\partial_X \tilde H(X)\tilde\psi\rangle\\
&\qquad +\langle \tilde\psi,\partial_X \tilde H(X)\tilde\psi\rangle.
\end{array}
\]
Use $\tilde H\Psi_0=0$ and consequently
\begin{equation}\label{H_der}
\partial_X\tilde H\Psi_0+ \tilde H\partial_X\Psi_0=0
\end{equation}
to obtain for the first term
\[
\begin{array}{rl}
 \langle \Psi_0,\partial_X \tilde H(X)\Psi_0\rangle &=
- \langle \Psi_0, \tilde H(X)\partial_X\Psi_0\rangle\\
&= -\langle \underbrace{\tilde H\Psi_0}_{=0}, \partial_X\Psi_0\rangle=0.\\
\end{array}
\]
Let $\Re$ denote the real part. The second terms are
\begin{equation}\label{second_split}
\begin{array}{rl}
\langle \tilde\psi,\partial_X \tilde H(X)\tilde\gamma_0\Psi_0\rangle
+\langle\tilde\gamma_0 \Psi_0,\partial_X \tilde H(X)\tilde\psi\rangle
&= 2\Re \langle \tilde\psi,\partial_X \tilde H(X)\tilde\gamma_0\Psi_0\rangle\\
&= -2\Re \langle \tilde\psi,\tilde H(X)\tilde\gamma_0\partial_X\Psi_0\rangle,\\
\end{array}
\end{equation}
and we have obtained the forcing
\begin{equation}\label{ehren_force}
\langle \bar\psi,\partial_X H(X)\bar\psi\rangle=\partial_X\lambda_0(X) 
-2\Re\langle\tilde\psi,\tilde H(X)\tilde\gamma_0\partial_X\Psi_0(X)\rangle + 
\langle\tilde\psi,\partial_X\tilde H(X)\tilde\psi\rangle.
\end{equation}
Lemma \ref{main_lemma} shows  that the third term, $\langle\tilde\psi,\partial_X\tilde H\tilde\psi\rangle$,  
is negligible  small at low temperature.

\subsection{Stochastic Electron Initial Data}\label{stok_init_sec}
The next step, to choose the stochastic initial data  $\tilde\psi^0$, 
applies the same reasoning as for the Zwanzig model in Sections \ref{zwanzig_initial} and \ref{gibbs_der}.
Similar to the partition of the Zwanzig-Hamiltonian in \eqref{heavy_light},
split the Ehrenfest-Hamiltonian in Lemma \ref{ehrenfest_zwanzig_lemma} into
\[
H_E= \underbrace{\frac{|p|^2}{2} +\lambda_0(X)}_{H_h} +\underbrace{\langle \tilde\psi,\tilde H\tilde\psi\rangle/\langle\psi,\psi\rangle}_{H_l}
\]
where the  "light" particle energy $H_l$ is associated to
the perturbation of the wave function
and  $H_h$ is the "heavy" nuclei energy. %
Here we used that the definition $\psi=\tilde\gamma_0\Psi_0 + \tilde\psi$ in \eqref{psi_tilde} and $\tilde H\Psi_0=0$ implies
$\langle\psi,\tilde H\psi\rangle/\langle\psi,\psi\rangle=\langle\tilde\psi,\tilde H\tilde\psi\rangle/\langle\psi,\psi\rangle$
and we also include the normalization factor  $1/\langle\psi,\psi\rangle$ since the normalization is an issue in determining the initial data.
Note that  it is the Hamiltonian 
$H_E= |p|^2/2 + \lambda_0(X)+ C\langle \tilde\psi,\tilde H(X)\tilde\psi\rangle$
that generates a Hamiltonian system, for any constant $C$, and not the normalized form $|p|^2/2 + \lambda_0(X) +\langle \tilde\psi,\tilde H(X)\tilde\psi\rangle/\langle\psi,\psi\rangle$; since  the Hamiltonian system conserves the $L^2(dx)$-norm $\langle\psi^t,\psi^t\rangle$
with respect to time, the normalization can be put in the equation for $p$ by taking $C=1/\langle \psi,\psi\rangle$,
to get the correct forces.
In this section we therefore first motivate the distribution of the non normalized initial $\tilde\psi^0$
and then when this distribution is chosen  we normalize each sample of initial wave function $\tilde\psi^0$ to
obtain our stochastic initial realizations. Consider the nuclei in their initial position $X^0$,
diagonalize the initial electron energy and make a cut-off allowing only the
first $J$ modes 
\begin{equation}\label{psi_init}
\begin{split}
\tilde\psi^0 &=\sum_{j=1}^J \gamma_j\Psi_j(X^0),\\
\psi^0&=\gamma_0\Psi_0(X^0)+\tilde\psi^0,\\
C\langle \psi^0,\tilde H(X^0)\psi^0\rangle  & = C\langle \tilde\psi^0,\tilde H(X^0)\tilde\psi^0\rangle =C\sum_{j=1}^J\tilde\lambda_j(X^0)|\gamma_j|^2,\\
\end{split}
\end{equation}
with the orthonormal eigenfunctions $\Psi_j(X^0)$ and eigenvalues $\tilde\lambda_j(X^0)$
to $\tilde H(X^0)$. 
We see that the initial light particle probability density satisfies
\[
e^{-C\langle\tilde\psi,\tilde H\tilde\psi\rangle/T} d\tilde\psi^r d\tilde\psi^i
=e^{-\sum_{j=1}^J \frac{C\tilde\lambda_j}{T} (|\gamma^j_r|^2 +|\gamma_i^j|^2)} d\gamma_1^r\ldots d\gamma_J^r 
d\gamma_1^i\ldots d\gamma_J^i,
\]
 since the orthogonal transformation from 
$(\tilde\psi^r,\tilde\psi^i)$ to $(\gamma_1^r,\ldots,\gamma_J^r,\gamma_1^i,\ldots, \gamma_J^i)$  has the Jacobian determinant equal to one, as in \eqref{jacobi}.
Consequently this Gibbs distribution generates independent normal distributed $\gamma_j, \ j=1,2,\ldots,J$ with independent real and imaginary parts. For technical reasons it is simpler to work with bounded variables in the analysis:
the proof only requires the right mean and covariance asymptotically. In addition
our setting with the nuclei energy $H_h$ dominating the electron energy $H_l$, formulated in condition \eqref{T_vilkor} below, will
imply that $C=1/\langle\psi,\psi\rangle$ is to high accuracy approximated by $1$. Therefore we choose  %
\begin{equation}\label{psi_data}\begin{split}
&\mbox{$\{\gamma_j\ |\ j=1,\ldots, J\}$ are independent, with independent real and imaginary parts}\\ 
&\mbox{which has mean zero and variance $T/\tilde\lambda_j(X^0)$}\\
&\mbox{and normal (or uniform) distribution density in $\big(-(\frac{3T}{\tilde\lambda_j(X^0)})^{1/2},
(\frac{3T}{\tilde\lambda_j(X^0)})^{1/2}\big)$}\\
&\mbox{and zero density outside this interval.}
\end{split}
\end{equation}
This means that the forces generated by $\psi^\tau=\tilde\psi^\tau+\gamma_0\Psi_0(X^\tau)$
should be normalized by dividing by $\sum_{j=0}^J|\gamma_j|^2=\langle\psi^0,\psi^0\rangle=
\langle\psi^\tau,\psi^\tau\rangle$.
The normalized  initial wave function becomes
\begin{equation}\label{wave_initial}
\sum_{j=0}^J\tilde\gamma_j\Psi_j(X^0):=\sum_{j=0}^J \frac{\gamma_j}{(\sum_{k=0}^J|\gamma_k|^2)^{1/2}}\Psi_j(X^0).
\end{equation}
We will assume that the ground state dominates, i.e. that %
\[ %
\frac{\sum_{j=1}^J|\gamma_j|^2}{|\gamma_0|^2} =o(1), \mbox{ as $M\rightarrow\infty$}
\] %
which by the bounded support in \eqref{psi_data} follows from the assumption %
\begin{equation}\label{T_vilkor}
\sum_{j=1}^J T/\tilde\lambda_j=o(1), \mbox{ as $M\rightarrow\infty$},
\end{equation}
so that in particular $\tilde\gamma_0\simeq 1$.

\subsection{The friction coefficient}
To derive the Langevin equation  the spectrum of $\tilde H$ will be used
and the main assumption is that the distribution of eigenvalues $\tilde\lambda_j$
approaches  a continuum  as $J\rightarrow\infty$, corresponding to the limit operator $\tilde H_\infty$.
  Assume that $\tilde H$ %
satisfies %
the following limit,  as in %
 \eqref{z:generalized} and \eqref{k_z},
\begin{equation}\label{K_definition}
\begin{split}
&K_{mn}({X^\tau }) \\
&:= \mathbb E_\gamma\big [2
\lim_{M,J\rightarrow\infty}
\int_0^{\tau M^{1/2}} 
{\Re\langle \tilde  S_{\hat \tau,\hat \tau-\hat \sigma} \partial_{X_n}\Psi_0(X^{\tau-\hat \sigma M^{-1/2}}),\tilde H(X^\tau)\partial_{X_m}\Psi_0(X^\tau)\rangle}
 d\hat \sigma | X^\tau].\\
\end{split}
\end{equation}
The limit in the right hand side can be split into the two steps
\[
\begin{split}
&%
\lim_{M,J\rightarrow\infty} 2\int_0^{M^{1/2}\tau} %
\Re\langle \tilde  S_{\hat \tau,\hat \tau-\hat \sigma} %
\partial_X\Psi_0(X^{ \tau-M^{-1/2}\hat \sigma }), 
 \tilde H\partial_X\Psi_0(X^\tau)\rangle %
 d\hat \sigma \\
& =2\int_0^\infty \lim_{M,J\rightarrow\infty} %
\Re\langle 
{1_{\hat \sigma \le M^{1/2}\tau}
\tilde H(X^\tau)\tilde  S_{\hat \tau,\hat \tau-\hat \sigma} %
\partial_X\Psi_0(X^{\tau-M^{-1/2}\hat \sigma })} ,
\partial_X\Psi_0(X^\tau)\rangle   \,  %
d\hat \sigma \\
& =2\int_0^\infty %
\Re \langle \tilde H_\infty(X^\tau) e^{i\hat\sigma\tilde H_\infty(X^\tau)} \partial_X\Psi_0(X^{\tau}), 
\partial_X\Psi_0(X^\tau)\rangle  %
d\hat \sigma \ .\\
\end{split}
\]
The first equality uses dominated convergence, as in \eqref{k_z}. The second equality uses the separation
of slow and fast time scales, as $M\rightarrow\infty$. This means that we assume  that the slow $X$-dynamics
satisfies \[
X^\tau- X^{\tau - M^{-1/2}\hat\sigma}=\int_0^{M^{-1/2}\hat\sigma} p(\tau -\sigma) d\sigma=\mathcal O(M^{-1/2})\]
 and that the fast dynamics for the electron wave function, based on 
the solution operator in the fast time scale $\tilde  S_{\hat \tau,\hat \tau-\hat \sigma}v=\varphi(\hat\sigma)$ defined by
\[
\begin{split}
i\dot\varphi(s) &= \tilde H(X^{\tau-M^{-1/2}s})\varphi(s)\\
\varphi(0)&= v=\partial_X\Psi_0(X^{\tau-M^{-1/2}\hat\sigma})\ ,
\end{split}
\] 
satisfies as $M,J\rightarrow\infty$
\[
\lim_{M,J\rightarrow\infty} \varphi(\hat\sigma)=e^{i\hat\sigma\tilde H_\infty(X^\tau)} v\ .
\]
The limit $K$ forms a $3N\times 3N$ matrix when it exists.
Such a limit might be achieved in a different way if 
the spectrum of $\tilde H(X)$ becomes sensitive towards small perturbations of $X$
and generates a random matrix $\tilde H_\infty$,
see Section \ref{limit_K}.

\subsection{Slow Nuclear Dynamics}\label{sec:dynamics_data}

The Ehrenfest dynamics \eqref{X_eq}, \eqref{psi_eq}
can be written as a Hamiltonian system in the slow time scale
\[
\begin{array}{rl}
\dot X^{\tau} &=p^{\tau}\\
\dot p^{\tau} &=  -\partial_X\lambda_0  
-\langle\phi^{\tau}, \partial_X \tilde H(X^{\tau})
\phi^{\tau}\rangle/\langle\phi,\phi\rangle\\
\frac{i}{M^{1/2}}\dot {\phi}^{\tau} &=   \tilde H(X^{\tau})\phi^{\tau},\\
\end{array}
\]
with the Hamiltonian
\[
\frac{|p|^2}{2} + C\langle\phi, \tilde H(X)\phi\rangle + \lambda_0(X),
\]
using the constant $C=M^{1/2}/2= 1/\langle \phi,\phi\rangle$
and letting $\psi^{\hat\tau}=\phi^\tau/\langle\phi^\tau,\phi^\tau\rangle^{1/2}$
for the normalized initial data $\psi^0=\sum_{n=0}^{ J} (\sum_{k=0}^J|\gamma_k|^2)^{-1}\gamma_n\Psi_n(X^0)$;
here we change to the simpler notation $(X^\tau,p^\tau)$ for position and momentum in the slow time scale,
although to be consistent with \eqref{X_eq}-\eqref{psi_eq} it should have been $(X^{\hat\tau},p^{\hat\tau})$.
The change to the  normalized variable $\psi$  leads to
\begin{equation}\label{X_slow_eq}
\begin{split} %
\frac{ dX^{\tau}}{d\tau} &=p^\tau,\\
\frac{dp^\tau}{d\tau} &=-\langle \psi^{\tau},\partial_X H(X^{\tau})\psi^{\tau}\rangle,\\
\frac{i}{M^{1/2}}\frac{d\psi^{\tau}}{d\tau} &= \tilde H(X^{\tau})\psi^{\tau}.
\end{split} %
\end{equation}

\subsection{The Main Result}
Let $W^\tau$ denote the standard Brownian process (at time $\tau$)
in $\rset^{3N}$ with independent components.
To simplify the  notation, %
assume that all nuclei have the same mass $M\gg 1=\mbox{electron mass}$; this can easily be extended to varying
nuclear masses much larger than the electron mass. 
We apply the notation $\psi(x,X)=\mathcal O(M^{-\alpha})$
also for complex valued functions, meaning that
$|\psi(x,X)|=\mathcal O(M^{-\alpha})$ holds uniformly in $x$ and $X$.

The approximation results
considers error estimates as $M\rightarrow\infty$ and for the assumptions to hold we also need $T, J$ and $ N$ to depend on $M$: assumption \eqref{K_definition} typically requires 
$J\rightarrow \infty$, as seen in the Zwanzig example \eqref{debye_fordel}-\eqref{delta},
and we assume that $M\rightarrow\infty$ fast compared to $J\rightarrow\infty$ so that \eqref{K_definition} holds (see also Section \ref{limit_K});
then $J\rightarrow\infty$   implies by  \eqref{T_vilkor}  that $T\rightarrow 0$ (slowly e.g.  as  $1/\log J$ explained  in \eqref{JT})
and finally \eqref{H_bound}  yields $N\rightarrow\infty$ to have $TJ/N\rightarrow 0$.
More precisely, we will use the following assumptions  for the electron eigenvalues $\tilde\lambda_j$ and electron Hamiltonian $\tilde H$
\begin{equation}\label{gap_villkor}
\begin{array}{rl}
T\sup_X\sum_{j=1}^J \frac{|\partial_X\tilde\lambda_j(X)|_{\ell^1(\rset^{3N})}}{\tilde\lambda_j(X)} & = o(M^{-1/2}),\\
 \sup_X\| |\tilde H \partial_X\Psi_0(X)|_{\ell^1(\rset^{3N})} \|_{L^2(dx)}&=\mathcal O(1),\\
\sup_{X} \|\partial_{X_k}\tilde H(X)\psi\|_{L^2(dx)} +\sup_{X} \|\partial_{X_jX_k}\tilde H(X)\psi\|_{L^2(dx)} &=\mathcal O(1),\\
X\mapsto \tilde H(X)& \mbox{ is smooth},\\
|\tilde\lambda_n-\tilde\lambda_m| >c \mbox{ for $n\ne m$ and } c^{-1}&=o(M^{1/5}),\\
\sum_{j=1}^JT\tilde\lambda_j^{-1} & = o(1).\\
\end{array}
\end{equation}
The conditions are motivated and used as follows.
The first {condition} in \eqref{gap_villkor}  means that the electron surfaces $\lambda_j$
are almost parallel to $\lambda_0$ and it
implies that the quadratic forcing term
$\langle \tilde\psi,\tilde H'\tilde\psi\rangle$ becomes negligible. The second and third condition are technical assumptions to obtain bounds in Lemmas \ref{main_lemma} and \ref{g_lemma}.
The fourth and fifth assumption (that the electron eigenvalues do not cross  along solution paths)
 are used in the Born-Oppenheimer approximation Lemma \ref{born_oppen_lemma}.
The last condition implies that the normalization of the wave function has a negligible effect on the distribution
in \eqref{psi_data}
and it also implies that the condition $\|H_l/H_h\|_{L^\infty}\ll 1$ in \eqref{H_bound} holds
so that sampling from the Gibbs distribution is reasonable,  in the sense of Section \ref{gibbs_der}.
If the density of states would be uniform and $\tilde\lambda_1=J^{-1}$, we would need $T= o (1/\log J)$ since  the last condition in \eqref{gap_villkor} yields
\begin{equation}\label{JT}
1\gg T\sum_{j=1}^J \tilde\lambda_j^{-1}\simeq T\int_{J^{-1}}^J \frac{d\lambda}{\lambda}=2T\log J,
\end{equation}
and in general we expect that the last condition in \eqref{gap_villkor} implies that $T\rightarrow 0$ as $J\rightarrow\infty$. %

As for the Zwanzig model studied in Refs.   \cite{stuart1,stuart2},   \cite{hald},  the following result
compares expected values of
Hamiltonian dynamics, having stochastic initial data $\gamma$ for the light particles,  with Langevin dynamics
where the stochasticity enters through the Wiener process $W$.
We use the notation $\mathbb E_z[w]$ for the expected value of $w$ with respect to the distribution generated by the random $z$,
the eigenvalue
$\lambda_0$ denotes the ground state electron energy \eqref{ground_state} of $H$ with normalized ground state $\Psi_0$
and $\tilde\lambda_j$ are the translated eigenvalues of $H-\lambda_0=:\tilde H$. 

\begin{theorem}\label{thm} Assume that condition \eqref{gap_villkor} 
and the friction limit \eqref{K_definition} hold,
the temperature $T$ is low enough to satisfy $T\sum_{j=1}^J\tilde\lambda_j^{-1}=o(1)$, and
 the Ehrenfest electron initial data is given by \eqref{psi_data}-\eqref{wave_initial}, %
then the It\^{o} Langevin dynamics
\begin{equation}\label{langevin}
\begin{array}{rl}
dp_L^\tau
&= -\partial_{ {X_L}} \lambda_0(X_L^\tau)
d\tau
-M^{-1/2}K(X_L^\tau)p^\tau_L d\tau
+\sqrt{2TM^{-1/2}}   K^{1/2}(X_L^\tau) dW^{\tau}, \\
\dot{X}_L^\tau &=p_L^\tau, \quad 0<\tau<\mathcal{T},
\end{array}
\end{equation}
with deterministic initial data $X_L^0=X$ and $p_L^0=p$
approximates Ehrenfest dynamics $(X^\tau,p^\tau)$ in \eqref{X_slow_eq}  with accuracy
\begin{equation}\label{rate}
\begin{split}
&\Big|
\mathbb{E}_\gamma\big[  g(X^{{\mathcal{T}}},p^{{\mathcal{T}}};X,p)\ \big | \ X^0=X,  p^0=p\big]\\
&\qquad-\mathbb{E}_W\big[ g( X_L^\mathcal{T}, p_L^\mathcal{T};X,p) \ \big | \ X_L^0=X,  p_L^0=p
\big]\Big|
=o(M^{-1/2}),
\end{split}\end{equation}
as $M\rightarrow\infty$, for any bounded function $g:\rset^{3N}\times\rset^{3N}\rightarrow \rset$, provided
the Langevin expected value 
\[
u(Y,q,\tau;\mathcal T):=\mathbb{E}_W[g(X_L^{\mathcal{T}},p_L^{\mathcal{T}}; X,p)\ |\ X_L^\tau=Y,\ p_L^\tau=q]
\]
has bounded derivatives of order one and two along the Ehrenfest solution
\begin{equation}\label{du_bound}
\begin{split}
&\int_0^{\mathcal{T}} 
\sum_j|\partial_{p_j}u(X^\sigma,p^\sigma,\sigma;\mathcal T)|  \, d\sigma=\mathcal{O}(1), \\
&\int_0^{\mathcal{T}} %
\sum_{j,k}\big(|\partial_{p_jp_k}u(X^\sigma,p^\sigma,\sigma;\mathcal T)|  
+|\partial_{X_jp_k}u(X^\sigma,p^\sigma,\sigma;\mathcal T)|\big) \, d\sigma=\mathcal{O}(1).
\end{split}
\end{equation}
\end{theorem}

The approximation result uses a non interacting particle, with given velocity equal to one and position coordinate 
$X_0=\tau$, that
acts as the time coordinate,
so that e.g. transport coefficients as diffusion $D$  can be studied
\[
\begin{split}
D &=\mathbb E_W[\underbrace{(6N\mathcal{T})^{-1}| X_L^{\mathcal{T}}-X_L^0|^2}_{=g(X_L^\mathcal{T}, \, 
p_L^\mathcal{T};X_L^0,p^0_L)}\ \big|\ X^0].\\
\end{split}
\] 
Since $(X_L^0,p^0_L)$ is deterministic and fixed, %
we simplify the notation and write $g(Y,q)$ instead of $g(Y,q;X_L^0,p^0_L)$.

\begin{remark}\label{observable}
It is well known that not all functions $g( X_L^\mathcal T,\dot{X}_L^\mathcal T)$ (such as $g(X,p)=\max_j |p_j|$)
are accurately computable in a molecular dynamics simulation with many particles; 
assumption \eqref{du_bound} restricts
the study to observables $g$ that are stable
with respect to perturbations in the initial data $(X_L^0,\dot{X}_L^0)$ in
the $\ell^1$ norm.
We consider expected values $E[g(X_L^{\mathcal T},p^\mathcal T_L)]$ where typically 
$g=\sum_{j=1}^N g_j/N$ is a mean over (particle related) real-valued functions $g_j$
with $g_j(X,p)=\mathcal{O}(1)$ and 
\[ 
\sum_{n=1}^N |\partial_{X_n} g(X,p)|
= \sum_{n=1}^N |\sum_{j=1}^N\frac{\partial_{X_n} g_j(X,p )}{N}|
=\mathcal{O}(1).\]
Here, for each derivative $\partial_{X_n}$, mainly finitely many $j$ contribute to the sum.
Similarly we assume that $ \sum_{n=1}^N |\partial_{X_n} u(X,p,\sigma)| =\mathcal{O}(1)$. 
This means that $g$ measures global properties, related to thermodynamic quantities, e.g. the diffusion
$D$ with $g=\sum_{j=1}^N |X_j-X_j^0|^2/(6N\mathcal T)$.
\end{remark}

Assumption \eqref{du_bound} is expected to hold for bounded times $\mathcal T=\mathcal O(1)$
but not for much longer times since the correlation time is expected to be long
as $M^{1/2}$ (so that  $\lim_{\tau\rightarrow\infty}\int_0^\tau 
\partial_p u(X^\sigma,p^\sigma,\sigma;\tau) d\sigma=\mathcal O(M^{1/2})$ ), explained as follows. Let $K$ be constant and positive,
then $\dot X_L=:p_L$ solves an
Ohrnstein-Uhlenbeck equation, which means that  
\[
\begin{split}
p^\tau_L&=(2TM^{-1/2}K)^{1/2}\int_0^\tau e^{M^{-1/2}(\sigma-\tau)K} dW^\sigma\\
&  - \int_0^\tau e^{M^{-1/2}(\sigma-\tau)K}  \lambda_0'(X^\sigma)\,  d\sigma 
+ e^{-M^{-1/2}K\tau}p^0
\end{split}
\]
and the dependence of $p^\tau$ on $p^\sigma$ decays exponentially $e^{M^{-1/2}(\sigma-\tau)K}$.
If also $\lambda_0'=0$, the process $p^\tau$ is Gaussian with mean zero and covariance 
\[
2Te^{-M^{-1/2}K|\tau-\sigma|} + e^{-M^{-1/2}K(\tau+\sigma)}(\mathbb E[|p^0|^2] - 2T)
\]
and
the corresponding $\partial_pu(\cdot,\sigma;\tau)$  
decays  exponentially $e^{-M^{-1/2}|\tau-\sigma|}$, %
giving the correlation length \[\int_0^\tau \partial_pu(\cdot, \sigma;\tau) d\sigma=\mathcal O(M^{1/2});\]
Section \ref{magnus_sec} presents a motivation for this property also in the case with a general force $\lambda_0'(X)$
and a constant positive friction matrix $M^{-1/2}K=kI$,
using the theory of large deviations for the rare events of escapes from equilibria at low temperature.

\begin{theorem}\label{long_time_thm}
Assume that  condition \eqref{du_bound} in Theorem \ref{thm} is replaced by
\begin{equation}\label{corr_bound}
\lim_{\tau\rightarrow\infty} \int_0^{\tau}  |\mathcal Du(X^\sigma,p^\sigma,\sigma;\tau)|_{\ell^1} \, d\sigma=\mathcal{O}(M^{1/2}), 
\quad \mathcal D:=\partial_p ,\partial_{pp},\partial_{Xp}, 
\end{equation}
i.e. the time-correlation length with respect to sensitivity in $p$ is at most of size $M^{1/2}$, 
then Langevin dynamics approximates long time observables of Ehrenfest dynamics 
\begin{equation}\label{rate_corr}
\lim_{\mathcal T\rightarrow\infty} \mathcal T^{-1}\big| \int_0^{\mathcal T}
\mathbb{E}_\gamma\big[  g(X^{{\tau}},p^{{\tau}})]
-\mathbb E_W\big[ g( X_L^\tau, p_L^\tau) %
\big] \ d\tau\big|
=o(1)\quad \mbox{ as $M\rightarrow\infty$}.
\end{equation}
\end{theorem}

The combination of
the two theorems  %
shows that this Langevin dynamics is in a sense an accurate approximation of Ehrenfest dynamics for 
both  short and long time. 
The small dissipation term $M^{-1/2}Kp_L$  in \eqref{langevin}
is visible in the convergence rate $o(M^{-1/2})$;
the small diffusion parameter  $TM^{-1/2}K=o(M^{-1/2})$
can be detected in the long  time convergence rate $o(1)$, since
the invariant measure is
\[
\frac{ e^{-(|p|^2/2+\lambda_0(X))/T} dX dp}{\int_{\rset^{3N+J}} e^{-(|p|^2/2+\lambda_0(X))/T} dX dp}
\]
and this holds if and only if  the diffusion coefficient is $(2TM^{-1/2}K)^{1/2}$, provided the dissipation  term 
is $M^{-1/2}Kp_L$.
That is,
the Langevin dynamics \eqref{langevin} satisfies   Einstein's fluctuation-dissipation result:
the square of the diffusion coefficient is the dissipation coefficient times twice the temperature,
as in \eqref{zwanzig_friktion}.

\begin{remark}
The Hamiltonian dynamics generated from the time-independent Schr\"odinger equation \eqref{schrod_t_i}, studied in Ref.   \cite{ASz}, is
closely related to the Ehrenfest dynamics where the Hamiltonian $H_E$ is essentially replaced
by  the slightly perturbed Hamiltonian $H_E + (2M)^{-1}G(X)\Delta_{X} (\langle \phi,\phi\rangle/G(X))$
for a certain bounded function $G(X)$, which is different for caustic and non caustic states. 
Therefore, it is possible that similar approximation results
hold when the Langevin dynamics is compared to this Schr\"odinger Hamiltonian dynamics.
\end{remark}

\subsection{Other Initial Electron Distributions}\label{entropy}

This section compares our model of initial data
with two other models of electron initial data, having given probabilities to be in mixed states or in pure eigenstates.
The theorems in this paper are based on
the Ehrenfest system viewed as a (classical) Hamiltonian system and sampling the electron configuration from its (classical) Gibbs distribution, conditioned on the nuclei positions, 
\[
\frac{e^{-\sum_{j=1}^J \frac{C\tilde\lambda_j}{T} (|\gamma^j_r|^2 +|\gamma_i^j|^2)} d\gamma_1^r\ldots d\gamma_J^r 
d\gamma_1^i\ldots d\gamma_J^i}{\int
e^{-\sum_{j=1}^J \frac{C\tilde\lambda_j}{T} (|\gamma^j_r|^2 +|\gamma_i^j|^2)} d\gamma_1^r\ldots d\gamma_J^r 
d\gamma_1^i\ldots d\gamma_J^i}\ .
\]
This turns out to be different from sampling the electron configurations in the distribution
where the probability to be in electron energy $\tilde\lambda_j$
is  ${e^{-\tilde\lambda_j/T}}/{\sum_je^{-\tilde\lambda_j/T}}$. For example, Einstein's fluctuation-dissipation
result, which holds for  the Ehrenfest Hamiltonian system model using its corresponding Gibbs distribution,  %
does not hold for the distribution where the probability to be in electron energy $\tilde\lambda_j$
is  ${e^{-\tilde\lambda_j/T}}/{\sum_je^{-\tilde\lambda_j/T}}$, since it generates a different covariance to \eqref{zwanzig_kovar} as explained below.

\subsubsection{Canonical Mixed States}
Let $q_j$ denote the density of state $j$ in the initial data $\psi^0$ composed of mixed states.
A usual setting of a canonical Gibbs-Boltzmann distribution is the choice 
$q_j={e^{-\tilde\lambda_j/T}}/{\sum_je^{-\tilde\lambda_j/T}}$,
which is motivated by maximizing the von Neumann entropy defined by $-\sum_j q_j\log q_j$, with the probability and energy constraints
$\sum_j q_j=1$ and $\sum_j\tilde\lambda_jq_j=\mbox{constant},$ see Ref.   \cite{gardiner}.
The stochastic model for the variable $|\gamma_j|^2$, giving in \eqref{psi_init} and \eqref{psi_data} the probability 
$|\gamma_j|^2/\sum_{k=0}^J |\gamma_k|^2$ to be in electron state $j$,
is  different from $q_j$. Namely, by \eqref{psi_data}-\eqref{T_vilkor} we have that
$\mathbb{E}[|\gamma_j|^2/\sum_{k=0}^J |\gamma_k|^2]$ 
is asymptotically  $T\tilde\lambda_j^{-1}$, for $j\ge 1$.
\subsubsection{Canonical Pure States}
Assume, instead of \eqref{psi_init}, that $\psi^0$ is a {\it pure electron eigenstate} $e^{i\alpha_j}\Psi_j$ with probability
\begin{equation}\label{quant_prob}
q_j:= e^{-\bar\lambda_j/T}(\sum_\ell e^{-\bar\lambda_\ell/T})^{-1}
\end{equation}
(and independent random phase shifts $\alpha_j$
uniformly distributed on $[0,2\pi]$) for { $j=0,\ldots, \tilde J$}, and write
$\psi^0 =:\sum_{j\ge 0}\check\gamma_j\Psi_j$
which has the covariance $\mathbb{E}[(\check\gamma_j)^*\check\gamma_k]= q_j\delta_{jk}$
and only one of the $\check\gamma_j,\ j=0,\ldots,\tilde J$, is non zero.
Let $\bar\Psi_0(X)=\check\gamma_0\, \Psi_0(X)$.
Then the fluctuations are very different from the case in Theorem \ref{thm},
since $\mathbb{E}[(\Re\langle \tilde \psi^0,\tilde H\partial_X\bar\Psi_0\rangle)^2]$ is zero
due to $\mathbb{E}[\check\gamma_j^*\check\gamma_k\ \check\gamma_0^*\check\gamma_0]=0$ for $j,k>0$.
Ehrenfest dynamics can approximate an eigenvalue problem for the full nuclei-electron system \eqref{schrod_t_i}, see Ref.   \cite{ASz};
that is different from \eqref{quant_prob} having the electrons in eigenstates of the electron operator $H$ for fixed nuclei positions.

\section{Error Estimates}\label{sec_approx}

\subsection{The Dynamics}
We shall approximate the nuclei motion in the Ehrenfest dynamics $(X^\tau,p^\tau)$  by $(X_L^\tau,p_L^\tau)$ defined from the Ito-Langevin dynamics
\begin{equation}\label{langevin_ekv}
\begin{array}{rl}
\dot X_L &=p_L\\
\dot p_L &=  -\partial_X\lambda_0(X_L)  - M^{-1/2} K(X_L) p_L + (2TM^{-1/2})^{1/2}  K^{1/2}(X_L) \dot W.\\
\end{array}
\end{equation}
To simplify the analysis of the coupling between $(X,p)$ and the thermal fluctuations induced by $\tilde\gamma_n$, introduce
the electron wave functions $\{\tilde\psi_n\ |\ n=0,1,\ldots \}$ which initially are normalized eigenfunctions
and solve %
\begin{equation}\label{psi_evolution}
\frac{i}{M^{1/2}}\dot {\tilde\psi}_n^\tau =   \tilde H( X^\tau)\tilde\psi_n^\tau, \quad \tilde\psi_n^0=\Psi_n(X^0),
\end{equation}
to obtain %
the normalized description
\begin{equation}\label{psi_definitionen}
\psi^{\hat\tau}=\sum_{n=0}^{ J} \frac{\gamma_n}{\big(\sum_{k=0}^J|\gamma_k|^2\big)^{1/2}}\tilde\psi_n
=:\sum_{n=0}^{ J}\tilde\gamma_n\tilde\psi_n
\end{equation}
where $\gamma_n$, defined in \eqref{psi_init},  are independent 
with mean zero and variance $T/\tilde\lambda_n(X^0)$ for $n=1,\ldots  J$.
Note that $\{\tilde\gamma_n\}_{n=1}^J$ depend on the initial position $X^0$ and the temperature $T$ but not on time $\tau$; the time dependence in $\psi$ is only through the time dependent functions $\tilde\psi_n$.
The Schr\"odinger dynamics \eqref{psi_evolution} shows that $\{\tilde\psi_n^\tau\ |\ n=0,1,\ldots \}$ forms an
orthonormal set
\begin{equation}\label{orto_set}
\begin{array}{rl}
\frac{d}{d\tau}\langle \tilde\psi_n,\tilde\psi_m\rangle&= \langle -iM^{1/2}\tilde H\tilde\psi_n,\tilde\psi_m\rangle
+\langle \tilde\psi_n,-iM^{1/2}\tilde H\tilde\psi_m\rangle\\
&=
\langle\tilde\psi_n, iM^{1/2}\tilde H\tilde\psi_m\rangle
+\langle \tilde\psi_n,-iM^{1/2}\tilde H\tilde\psi_m\rangle=0,
\end{array}
\end{equation}
since the initial data $\{\tilde\psi_n^0\ |\ n=0,1,\ldots \}$  is orthonormal.

\subsection{Proof of Theorem \ref{thm}}\label{sec:representation}
Define for the given observable $g:\rset^{3N}\times\rset^{3N}\rightarrow\rset$ and the Langevin dynamics $(X_L^\tau,p_L^\tau)$  in \eqref{langevin_ekv}
the expected value 
\begin{equation}\label{u_definition}
u(y,\tau)
:=\mathbb{E}_W[ g( X_L^\mathcal T,p_L^\mathcal T) \ |\ (X_L^\tau,p_L^\tau)=y ],\end{equation}
which solves the Kolmogorov backward equation   %
\begin{equation}\label{kolmogorov_ekv}
\begin{split}
\partial_\tau u + p\cdot\partial_{X} u - \big(\partial_{X}\lambda_0(X)+M^{-1/2}K(X)p\big)\cdot\partial_{p} u\ +& \\
\qquad  + \sum_{m=1}^{3N}\sum_{n=1}^{3N}
M^{-1/2}TK_{mn}(X)\partial_{p_mp_n} u=0&\ ,\ \ \tau<\mathcal T\\
u(\cdot,\mathcal T)=g&.
\end{split}
\end{equation}
The  goal is to analyze the error 
$\mathbb{E}_\gamma[g(  X^\mathcal T,p^\mathcal T)]  - \mathbb E_W[g(X_L^\mathcal T,p_L^\mathcal T)]$ %
of the heavy nuclei in the Ehrenfest dynamics
approximated by the Langevin dynamics, for given deterministic initial data $X^0=X_L^0, p^0=p_L^0$.
This error can be written as the difference of the expected value of the
Langevin Kolmogorov solution evaluated at the final and initial Ehrenfest position-momentum point,
using  the definitions \eqref{u_definition} and $u(\cdot,\mathcal T)= g$ in \eqref{kolmogorov_ekv}:
\[
\begin{split}
&\mathbb{E}_\gamma[g(  X^\mathcal T,p^\mathcal T) \ |\ X^0=X,p^0=p] - \mathbb E_W[
g(X_L^\mathcal T,p_L^\mathcal T)  \ |\ X_L^0=X,p_L^0=p] \\
&= \mathbb{E}_\gamma[u(X^\mathcal T,p^\mathcal T,\mathcal T) \ |\ X^0=X,p^0=p] -u(X,p,0)\\
&=\mathbb{E}_\gamma[u(X^\mathcal T,p^\mathcal T,\mathcal T) -u(X^0,p^0,0) \ |\ X^0=X,p^0=p]. \\
\end{split} %
\]
The expected value in right hand side
will be written as an integral over time where the assumption \eqref{du_bound} makes the integral bounded.
Since all remaining expected values are with respect to the initial data $\gamma$ we
simplify by writing $\mathbb E=\mathbb E_\gamma$.
Telescoping cancelation, the Kolmogorov equation and the nuclei forces 
\[
\dot p%
=-\partial_X\lambda_0(X) 
+2\Re\langle\tilde\psi,\tilde H(X)\tilde\gamma_0\partial_X\Psi_0(X)\rangle - 
\langle\tilde\psi,\partial_X\tilde H(X)\tilde\psi\rangle\]
 in \eqref{ehren_force} imply that the approximation error
 can be written as an integral of the
residual of the Langevin Kolmogorov solution \eqref{kolmogorov_ekv} along the  Ehrenfest dynamics  
\begin{equation}\label{main_estimate}
\begin{split}
&\mathbb{E}[u(X^\mathcal T,p^\mathcal T,\mathcal T) -u(X^0,p^0,0)\ |\ X^0, p^0]  =\int_0^\mathcal T  \mathbb{E}[ \ du(X^\tau ,p^\tau ,\tau) \ |\ X^0, p^0] \\
&= \int_0^\mathcal T  \mathbb{E}[ \partial_\tau u(X^\tau ,p^\tau ,\tau) + \dot X^\tau \cdot \partial_Xu (X^\tau ,p^\tau ,\tau) + 
\dot p^\tau \cdot
\partial_p u(X^\tau ,p^\tau ,\tau) \ |\ X^0, p^0] \ d\tau \\
&= \int_0^\mathcal T  \mathbb{E}\Big[\underbrace{(\dot X^\tau - p^\tau )}_{=0}\cdot \partial_Xu (X^\tau ,p^\tau ,\tau)  \\
&\qquad\qquad + \big(\dot p^\tau  +\lambda_0'(X^\tau ) + M^{-1/2}K(X^\tau )p^\tau \big) \cdot
\partial_p u(X^\tau ,p^\tau ,\tau) \\
&\qquad\qquad  -TM^{-1/2}K(X^\tau )\partial_{pp} u(X^\tau ,p^\tau ,\tau) \ |\ X^0, p^0\Big] \ d\tau  \\
&= \int_0^\mathcal T  \mathbb{E}\Big[
\big( 2\Re\langle\tilde\psi,\tilde\gamma_0\tilde H\Psi_0'\rangle -\langle \tilde\psi,\tilde H' \tilde\psi\rangle + M^{-1/2}K(X^\tau )p^\tau \big) \cdot
\partial_p u(X^\tau ,p^\tau ,\tau) \\
&\qquad\qquad  -TM^{-1/2}K(X^\tau )\partial_{pp} u(X^\tau ,p^\tau ,\tau) \ |\ X^0, p^0\Big] \ d\tau. \\
\end{split}
\end{equation} %
Consider the solution $(X,p,\psi)$ to the Ehrenfest dynamics 
\eqref{X_slow_eq} and define for the  Schr\"odinger equation 
\[
\begin{split}
i \dot \varphi^\tau &=M^{1/2}\tilde H(X^\tau )\varphi^\tau , \quad \tau\ge \sigma\\
\varphi^\sigma &= w
\end{split}
\]
the solution operator in the slow time scale
\[
S_{\tau,\sigma}w := \varphi^\tau .
\]
The definition $\psi=\tilde\psi +\tilde\gamma_0 
\Psi_0$  %
yields the representation
\begin{equation}\label{sol_repr}
\begin{split}
\tilde\psi^\tau &= -
\tilde\gamma_0
\int_0^\tau S_{\tau,\sigma} \dot \Psi_0(X^\sigma) d\sigma +S_{\tau,0}\tilde\psi^0\\
&= -\tilde\gamma_0
\int_0^\tau S_{\tau,\sigma} \dot \Psi_0(X^\sigma) d\sigma 
+ \sum_{n=1}^J\tilde\gamma_n \tilde\psi_n^\tau
\end{split}
\end{equation}
which implies
\[
\begin{split}
\Re\langle\tilde\psi^\tau,\tilde\gamma_0\tilde H\Psi_0'(X^\tau)\rangle &=-|\tilde\gamma_0|^2
\Re\langle \int_0^\tau S_{\tau,\sigma} \dot \Psi_0(X^\sigma) d\sigma,\tilde H\Psi_0'(X^\tau)\rangle\\
&\qquad + \Re\langle \sum_{n=1}^J\tilde\gamma_n\tilde\psi_n^\tau,\tilde\gamma_0\tilde H\Psi_0'(X^\tau)\rangle .\\
\end{split}
\]
\begin{lemma}\label{main_lemma}
There holds
\begin{equation}\label{main_lemma_est}
\begin{split}
&\lim_{M\rightarrow\infty}
-M^{1/2}\mathbb E\big[2|\tilde\gamma_0|^2 %
\Re\langle \int_0^\tau S_{\tau,\sigma} \dot \Psi_0(X^\sigma) d\sigma,\tilde H\Psi'_0(X^\tau)\rangle\ | \ X^\tau,p^\tau\big]
=K(X^\tau)p^\tau,\\ %
&\lim_{M\rightarrow\infty}
M^{1/2}\mathbb E\big[2\Re\langle \sum_{n=1}^J\tilde\gamma_n\tilde\psi_n^\tau,\tilde\gamma_0\tilde H\Psi'_0(X^\tau)\rangle\cdot\partial_pu(X^\tau,p^\tau,\tau)\ | \ X^\tau,p^\tau\big]\\
&\qquad\qquad\qquad =TK(X^\tau)\partial_{pp}u(X^\tau,p^\tau,\tau),\\ %
&\lim_{M\rightarrow\infty} M^{1/2}\mathbb E\big[\langle \tilde\psi^\tau,\tilde H'\tilde\psi^\tau
\rangle\cdot\partial_pu(X^\tau,p^\tau,\tau)\ |\ X^\tau,p^\tau\big]=0.\\ %
\end{split}
\end{equation}
\end{lemma}

The lemma and \eqref{main_estimate}  imply
\[
\mathbb{E}[u(X^\mathcal T,p^\mathcal T,\mathcal T) -u(X^0,p^0,0)\ |\ X^0=X, p^0=p] =o(M^{-1/2})
\]
which proves Theorem \ref{thm}.

{\bf Proof of the Lemma.}
{\it The first limit: the friction term.}  %
We have 
\[
|\tilde\gamma_0|^2=
\frac{|\gamma_0|^2}{|\gamma_0|^2+\sum_{k=1}^J|\gamma_k|^2}= 1 +\mathcal O( \frac{\sum_{k=1}^J|\gamma_k|^2}{|\gamma_0|^2}). 
\]
The last condition in \eqref{gap_villkor} implies %
\begin{equation}\label{last_cond}
 \frac{\sum_{k=1}^J|\gamma_k|^2}{|\gamma_0|^2}=o(1).
\end{equation}
The first statement follows then directly from the definition of $K$ in \eqref{K_definition},
 by the change of variables $M^{1/2}(\tau-\sigma)=\hat \sigma$ and integration of solution operator 
 $\tilde S_{\hat \tau,\hat \sigma}=S_{\tau,\sigma}$ in the fast time scale
 \[
 \begin{split}
& -M^{1/2}\Re\langle \int_0^\tau S_{\tau,\sigma} \dot \Psi_0(X^\sigma) d\sigma,
 \tilde H\Psi_0'\rangle \\
 &\qquad =
 \int_0^{\tau M^{1/2}} 
{\Re\langle \tilde  S_{\hat \tau,\hat \tau-\hat \sigma}\Psi_0'(X^{\tau-\hat \sigma M^{-1/2}}), \tilde H\Psi_0'(X^\tau)\rangle}
 d\hat \sigma.
\end{split}
 \] %

{\it The second limit: the diffusion term.} %
Define the first variation \[
\partial_{\tilde\gamma_n}( X, p, \psi)=:(X,p,\psi)'_n, \ n>0,\] which
satisfies the linearized Ehrenfest system
\begin{equation}\label{X'}
\begin{array}{rl}
\frac{d}{d\tau}\big((X^\tau_m)_n'\big) &= (p^\tau_m)'_n,\\
\frac{d}{d\tau}\big((p_m^\tau)'_n\big) &= -\sum_{k}(X_k^\tau)'_n\cdot\partial_{X_kX_m}\lambda_0( X^\tau)
- 2\Re \langle\tilde\psi_n^\tau, 
\partial_{X_m}\tilde H( X^\tau)\psi^\tau\rangle\\
&\quad - \langle \psi^\tau,\sum_{k}(X^\tau_k)'_n\cdot\partial_{X_kX_m}\tilde H(X^\tau)\psi^\tau\rangle\\
&\quad-2\Re\langle \sum_m\tilde\gamma_m(\tilde\psi_m^\tau)'_n,\partial_{X_m}\tilde H(X^\tau)\psi^\tau\rangle,\\
i\frac{d}{d\tau}\big((\tilde\psi_m^\tau)'_n\big)&=M^{1/2}\tilde H(X^\tau)(\tilde\psi_m^\tau)'_n + M^{1/2}\sum_{k}(X_k^\tau)'_n\cdot\partial_{X_k}\tilde H(X^\tau)\tilde\psi_m^\tau, \\
(X,p,\psi)'(0) &=0,\\
\end{array}
\end{equation}
where $\psi=\sum_{m=0}^J \tilde\gamma_m\tilde\psi_m$.
Duhamel's representation shows that 
\begin{equation}\label{koppling}
\begin{array}{rl}
(X,p,\psi)'_n(\tau) &=-2\int_0^\tau G_{\cdot p}(\tau,\sigma)\Re \langle\tilde\psi_n, 
\partial_X\tilde H( X)\psi\rangle(\sigma) d\sigma\\
&\simeq
-2\int_0^\tau G_{\cdot p}(\tau,\sigma)\Re\langle\tilde\psi_n^\sigma,\tilde H\partial_X\Psi_0^\sigma\rangle d\sigma
\end{array}
\end{equation}
and the last  step follows from
the asymptotic result $\psi\simeq \tilde\gamma_0\tilde\psi_0\simeq \Psi_0$, obtained from \eqref{last_cond} and
the Born-Oppenheimer approximation $\tilde\psi_0\simeq \Psi_0$  in Lemma \ref{born_oppen_lemma};
here $G$ is the linear solution operator for \eqref{X'}
solving 
\[
\mbox{$\dot G=AG, \ \tau>\sigma$ and $G(\sigma,\sigma)=I$;}\]
where
\[
A=A(X^\tau,\psi^\tau) := \left[\begin{array}{ccc}
0 & \quad\quad I & \quad\quad 0\\
-\lambda_0'' -\langle \psi,\tilde H''\psi\rangle& \quad\quad 0 & \quad\quad -2\Re\langle(\tilde\gamma_1,\ldots,\tilde\gamma_J),\tilde H'\psi\rangle\\
-iM^{1/2}\tilde H' B & \quad\quad 0 & \quad\quad -iM^{1/2}\tilde H I
\end{array}\right]
\]
 is the matrix in the right hand side of \eqref{X'} and
 the $J\times J$ matrix $B$ is diagonal with $\tilde\psi_j,\ j=1,\ldots J $ in the diagonal.
We need information about 
the short time behavior of $G(\tau,\sigma)$ as $\sigma\rightarrow \tau$. 
For short time the Green's function takes the form
$G(\tau,\sigma)=e^{(\tau-\sigma)A}$
and we have %
\begin{lemma}\label{g_lemma} There holds
\begin{equation}\label{f_pp_est}
\begin{split}
&G_{pp}(\tau,\sigma) \rightarrow I \quad \mbox{ as } \sigma\rightarrow \tau,\\
&{G_{Xp}(\tau,\sigma)} =\mathcal O(\tau-\sigma) \mbox{ for }   0\le \tau-\sigma=o(M^{-1/4}),\\
&\partial_{\tilde\gamma_n}\tilde\psi_m^\tau
=\\
&=-\sum_k iM^{1/2}\int_0^\tau
\underbrace{\int_\varsigma^\tau \int_\varsigma^\sigma S_{\tau,\sigma}G_{pp}(\upsilon,\varsigma) d\upsilon\  \partial_{X_k}\tilde H(X^\sigma)\tilde\psi_m^\sigma d\sigma}_{L^2(dx)\mbox{-norm}=\mathcal O((\tau-\varsigma)^2)} \langle \tilde\psi_n^\varsigma, \tilde H\partial_{X_k}\Psi_0(X^\varsigma)\rangle d\varsigma\, ,
\end{split}
\end{equation}
\end{lemma}
The proof is in the end of this section. We will now use the first variation 
with respect to $\tilde\gamma_n$ in \eqref{koppling}
to verify the second limit in \eqref{main_lemma_est}.
We know that
\[
\begin{split}
\mathbb E[\tilde\gamma_n] =\mathbb E[\frac{\gamma_n}{(|\gamma_0|^2 + \sum_{j=1}^J|\gamma_j|^2)^{1/2}}]
=0,
\end{split}
\]
since $\gamma_n$ is symmetric distributed with mean zero. Consequently,
if $(X,p,\psi)$ would be independent of $\tilde\gamma_n$ the expected value
\[
M^{1/2}\mathbb{E}[2\Re\langle \tilde\gamma_n\tilde\psi_n^\tau,\tilde\gamma_0\tilde H\Psi_0'(X^\tau)\rangle\cdot\partial_p u(X^\tau,p^\tau,\tau)]
\]
would be zero. The first variation with respect to $\tilde\gamma_n$ tells us  how $(X,p,\psi)$ depends on $\tilde\gamma_n$.
The coupling 
can be split into two terms -- one term that considers the coupling between the two factors
$\langle \tilde\gamma_n\tilde\psi_n^\tau ,\tilde\gamma_0\tilde H\Psi_0'(X^\tau )\rangle$ and $\partial_p u(X^\tau ,p^\tau ,\tau)$
and one term with the intrinsic coupling in the first factor 
$\langle \tilde\gamma_n\tilde\psi_n^\tau ,\tilde\gamma_0\tilde H\Psi_0'(X^\tau )\rangle$.
The coupling between the two factors is
\begin{equation}\label{u_koppling}
\begin{split}
 &\sum_{m=1}^J M^{1/2}\int_0^\tau\int_\rset \mathbb 
E\Big[2\Re\langle \tilde\gamma_n\tilde\psi_n^\tau  ,\tilde\gamma_0\tilde H\Psi_0'(X^\tau  )\rangle
2\Re\langle \tilde\psi_m^\sigma ,\tilde H'\psi^\sigma \rangle \times\\
& 
\big( G_{pp}(\tau,\sigma )\partial_{pp} u(X^\tau  ,p^\tau  ,\tau) + G_{Xp}(t,\sigma )\partial_{Xp}u(X^\tau  ,p^\tau  ,\tau)\big)
 \Big| \underbrace{\tilde\gamma_m^r + i\tilde\gamma_m^i}_{=\tilde\gamma_m}\Big] (d\tilde\gamma_m^r + id\tilde\gamma_m^i) d\sigma.\\ 
  \end{split}
\end{equation}
The intrinsic coupling is equal to
\begin{equation}\label{intrinsic}
\begin{split}
&\sum_{m=1}^J M^{1/2}\int_0^\tau  \int_\rset \mathbb 
E\Big[
2\Re\langle \tilde\gamma_n\tilde\psi_n^\tau,\tilde\gamma_0\partial_X(\tilde H\Psi_0')G_{Xp}(\tau,\sigma)\rangle \\
&\times2\Re\langle \tilde\psi_m^\sigma ,\tilde H'\psi^\sigma \rangle 
\partial_pu(X^\tau  ,p^\tau  ,\tau)
  \ \Big|\ \tilde\gamma_m^r + i\tilde\gamma_m^i\Big] (d\tilde\gamma_m^r + id\tilde\gamma_m^i) d\sigma\\
  &+\sum_{m=1}^J M^{1/2}\int_\rset
  \mathbb{E}\big[2\Re\langle \tilde\gamma_n\partial_{\tilde\gamma_m}\tilde\psi_n^\tau  ,\tilde\gamma_0\tilde H\Psi_0'(X^\tau  )\rangle\cdot\partial_p u(X^\tau  ,p^\tau  ,\tau)  \ \big|\ \tilde\gamma_m^r + i\tilde\gamma_m^i\big] 
  d\tilde\gamma_m
  \end{split}
\end{equation} %
where the first variation $\partial_{\tilde\gamma_m}\tilde\psi_n$ is expressed in terms of $G$ in \eqref{f_pp_est}.

Let us first study the coupling \eqref{u_koppling} between the two factors. The forcing
has the asymptotics
\[
\begin{split}
\tilde H'\psi&= \sum_{n=0}^J\tilde\gamma_n\tilde H'\tilde\psi_n\\
&\simeq \tilde\gamma_0\tilde H'\tilde\psi_0\\
&\simeq \tilde\gamma_0\tilde H'\Psi_0\\
&=-\tilde H\Psi_0'
\end{split}
\]
since $\sum_{n=1}^J\tilde\gamma_n\tilde\psi_n$ is asymptotically
smaller than $\tilde\gamma_0\tilde\psi_0$ in $L^2(dx)$, by the last condition in \eqref{gap_villkor}, and $\tilde\psi_0\simeq \Psi_0$
by Lemma \ref{born_oppen_lemma}.
 This simplified
forcing is used to evaluate to coupling below.
The first variation shows that the first term in the perturbation in $p$ due to $\gamma_m$ becomes
\begin{equation}\label{G_p_pert}
\begin{split}
&\int_0^\tau \int_0^{\tilde\gamma_m} \underbrace{\big(G_{pp}(\tau,\sigma)\partial_{pp}u +G_{Xp}(\tau,\sigma)\partial_p u\big)
\langle \tilde\psi_n^\sigma,\tilde H\Psi_0'(X^\sigma)\rangle \langle\tilde H\Psi_0'(X^\tau), \tilde\psi_m^\tau\rangle }_{
=:F(\hat\gamma_m)}\\
&\qquad\times
|\tilde\gamma_0|^2\tilde\gamma_n^* d\hat \gamma_m \ d\sigma.
\end{split}
\end{equation}
The last condition in \eqref{gap_villkor} implies that
$|\tilde\gamma_0|^2 \simeq 1$, so that
\[
|\tilde\gamma_0|^2\tilde \gamma_n^* d\hat\gamma_n
\simeq  \gamma_n^* d\hat\gamma_n
\]
and the expected perturbation,  for $m=n$,  can  be written 
\begin{equation}\label{F_est}
\begin{split}
&\mathbb E[ \int_0^\tau \int_0^{\tilde\gamma_n} F(\hat\gamma_n)
|\tilde\gamma_0|^2\tilde \gamma_n^r d\hat\gamma_n^r
\ d\sigma]\\
&\simeq \int_0^\tau  \int_\rset F(0)|\gamma_n^r|^2\,  \rho_n(\gamma_n^r) d\gamma_n^r\, d\sigma\\
&=  \int_0^\tau  F(0)
\frac{T}{\tilde\lambda_n}\ d\sigma,
\end{split}
\end{equation}
using the density $\rho_n$ of $\gamma_n$ with the mean zero and variance $T/\tilde\lambda_n$ given in \eqref{psi_data}.
The imaginary part $\gamma_n^i$ gives an identical contribution.
To see that that the approximation $F(\tilde\gamma_n)\simeq F(0)$ holds,
we can similarly use the coupling applied to $F$ to find its dependence on $\tilde\gamma_n$
\[
\begin{split}
F(\tilde\gamma_n) &= F(0) +
\int_0^\tau \int_0^{\tilde\gamma_n} G_{pp}(\tau,\sigma)\partial_p F
 \langle\tilde H\Psi_0'(X^\tau), \tilde\psi_n^\tau\rangle \gamma_0 d \hat\gamma_n \, d\sigma
 +\ldots.\\
\end{split}
\]
We see that the dependence on $\tilde\gamma_n$  is small and proportional to %
\[
\langle \tilde H\Psi_0', \tilde\psi_n \rangle=\mathcal O(\sqrt{\Delta\lambda})=o(1)\]
where $\Delta\lambda$ is the  distance between the eigenvalues.
Therefore the perturbation \eqref{G_p_pert} in $F$ due to $\tilde\gamma_n$ is small
\[
F(\tilde\gamma_n)=F(0) + \mathcal O((\Delta\lambda)^{1/2})\simeq F(0).
\]
The expected value of the coupling between $\tilde\gamma_n$ and $\tilde\gamma_m$, for $m\ne n$,
is to leading order equal to zero, since 
\[
\int_\rset  F(0) \gamma_n^*\gamma_m\rho_m(\gamma_m)d\gamma_m
=0\, .
\]
The first non zero expected value for $n\ne m$ can again be obtained by
expanding $F$ with respect to both $\tilde\gamma_m$ and $\tilde\gamma_n$
\begin{equation}\label{off_diagonal}
\begin{split}
F(\tilde\gamma_m) &= F(0) +
\int_0^\tau \int_0^{\tilde\gamma_m} G_{pp}(\tau,\sigma)\partial_p F
 \langle\tilde H\Psi_0'(X^\tau), \tilde\psi_m^\tau\rangle \tilde\gamma_0 d \hat\gamma_m \, d\sigma
 +\ldots\\
& =F(0) +\int_0^\tau \int_0^{\tilde\gamma_m} \int_0^\tau \int_0^{\tilde\gamma_n}
 G_{pp}(\tau,\sigma)\partial_{pp} F
 \langle\tilde H\Psi_0'(X^\tau), \tilde\psi_m^\tau\rangle \tilde\gamma_0 \\
&\qquad\qquad\qquad\qquad\qquad \times G_{pp}(\tau,\varsigma)
 \langle\tilde H\Psi_0'(X^\tau), \tilde\psi_n^\tau\rangle \tilde\gamma_0 d \hat\gamma_n \, d\varsigma 
 d \hat\gamma_m \, d\sigma
\end{split}
\end{equation}
and it takes a similar form as  the coupling with only one factor $\tilde\gamma_n$
but now a product of two such coupling factors appear. We will see below that the each integral 
gains a factor of $M^{-1/2}$
\begin{equation}\label{F_uppskatt}
\int_0^\tau F(0) \frac{T}{\tilde\lambda_n} d\sigma  =\mathcal O(M^{-1/2}),
\end{equation}
so that the quadratic term with
both factors $\tilde\gamma_n$ and $\tilde\gamma_m$ is negligible small $\mathcal O(M^{-1})$.

The function $F$ contains the product
\begin{equation}\label{prod_four}
\big(\langle\ldots^\tau_n\rangle +\langle\ldots^\tau_n\rangle^*\big)
\big(\langle\ldots^\sigma_m\rangle +\langle\ldots^\sigma_m\rangle^*\big)
\end{equation}
and can be analyzed by the four terms $\langle\ldots^\tau_n\rangle\langle\ldots^\sigma_m\rangle$
similar as in \eqref{zwanzig_kovar}, now using the solution operator $S_{\tau\sigma}$ instead of the
explicit solution $e^{-iM^{1/2}(\tau-\sigma)\tilde H}$. %
The following five steps show that
the expected value of the fluctuations
takes the same form as the friction term in the first statement of  \eqref{main_lemma_est}:
\begin{itemize}
\item[1.] in the first step the error term comes from  $(X,p)$ being  slightly dependent on 
$\tilde\gamma_n$ -- this coupling
yields a small error term as estimated in \eqref{F_est}, 
\item[2.] the second step uses the first  condition in \eqref{gap_villkor} to deduce
that  $1/\tilde\lambda_n^0$ and $1/\tilde\lambda_n^\tau$ are close, as explained in \eqref{lambda_prim}-\eqref{gap_variation},
\item[3.] the third
step uses that $\tilde\psi_n=\Psi_n+o(1)$,  as derived in Lemma \ref{born_oppen_lemma},
to replace a factor of $\tilde\lambda_n^{-1}$ with $\tilde H^{-1}$,
\item[4.] the fourth step applies  $ J\rightarrow\infty$, that
$\langle (S_{\tau\sigma})^* \tilde H\Psi_0'(X^\tau),\Psi_0'(X^\sigma)\rangle$ is finite, and that
the orthonormal  set $\{\tilde\psi_n\}_{n=1}^\infty$ in \eqref{orto_set} forms a basis in the $L^2(\rset^J)$ orthogonal complement of
$\tilde\psi_0$, and 
\item[5.] the fifth step uses that $\Psi_0$ is orthogonal to $\partial_X\Psi_0$ and
$\tilde\psi_0\simeq\Psi_0$, proved in Lemma \ref{born_oppen_lemma},
\end{itemize}
we write the four terms formed from \eqref{prod_four} as the sum of two real parts as follows
\begin{equation}\label{fluct_diss}
\begin{array}{rl}
&2\Re\, \mathbb E[ \sum_{n=1}^{J}\langle  S_{\tau,\sigma}\tilde\gamma_n\tilde\psi_n^\sigma,\tilde H \partial_X\Psi_0^\tau\rangle 
\langle \tilde H \partial_X\Psi_0^\sigma,\tilde\gamma_n\tilde \psi_n^\sigma\rangle\ |\ X^\tau ]\\
&\qquad +2\Re\, \mathbb E[ \sum_{n=1}^{J}\langle 
 S_{\tau,\sigma}\tilde\gamma_n\tilde\psi_n^\sigma,
 \tilde H \partial_X\Psi_0^\tau
 \rangle 
\langle\tilde\gamma_n\tilde \psi_n^\sigma, \tilde H \partial_X\Psi_0^\sigma\rangle\ |\ X^\tau ]\\
&  \underbrace{\simeq}_{1.} 2 \Re \mathbb E\big[\sum_{n=1}^{ J}\langle  S_{\tau,\sigma}\tilde\psi_n^\sigma,\tilde H \partial_X\Psi_0^\tau\rangle 
\langle \tilde H \partial_X\Psi_0^\sigma,\tilde \psi_n^\sigma\rangle \underbrace{\mathbb{E}[\tilde\gamma_n^*\tilde\gamma_n]\ |\ X^\tau\big]}_{\simeq 2T(\tilde\lambda_n^0)^{-1}}\\
&\qquad 
+ 2 \Re \mathbb E\big[\sum_{n=1}^{ J}\langle S_{\tau,\sigma}\tilde\psi_n^\sigma,\tilde H \partial_X\Psi_0^\tau\rangle 
\langle \tilde \psi_n^\sigma,\tilde H \partial_X\Psi_0^\sigma\rangle \ |\ X^\tau\big]
\underbrace{\mathbb{E}[\tilde\gamma_n^*\tilde\gamma_n^*]}_{=0}\\
&= 4 \Re \mathbb E\big[\sum_{n=1}^{J} \langle S_{\tau,\sigma}\tilde\psi_n^\sigma,\tilde H \partial_X\Psi_0^\tau\rangle 
\langle \tilde H \partial_X\Psi_0^\sigma,\tilde \psi_n^\sigma\rangle \frac{T}{\tilde \lambda_n^0}\ |\ X^\tau\big]\\
&\simeq 4 \Re \mathbb E\big[\sum_{n=1}^{ J} \langle S_{\tau,\sigma}\tilde\psi_n^\sigma,\tilde H \partial_X\Psi_0^\tau\rangle 
\langle \partial_X\Psi_0^\sigma, \tilde H\tilde \psi_n^\sigma\rangle \frac{T}{\tilde \lambda_n^\sigma}\ |\ X^\tau\big]\\
&\simeq 4 T\Re \mathbb E\big[
\sum_{n=1}^{ J} \langle S_{\tau,\sigma}\tilde\psi_n^\sigma,\tilde H \partial_X\Psi_0^\tau\rangle 
\langle  \partial_X\Psi_0^\sigma,\tilde \psi_n^\sigma\rangle \ |\ X^\tau\big]\\
& \simeq 4 T\Re \mathbb E\big[\sum_{n= 1}^\infty
 \langle \tilde\psi_n^\sigma,(S_{\tau,\sigma})^*\tilde H \partial_X\Psi_0^\tau\rangle 
\langle  \partial_X\Psi_0^\sigma,\tilde \psi_n^\sigma\rangle\ |\ X^\tau\big] \\
& \underbrace{\simeq}_{5.} 4 T\Re \mathbb E\big[\sum_{n= 0}^\infty
 \langle \tilde\psi_n^\sigma,(S_{\tau,\sigma})^*\tilde H \partial_X\Psi_0^\tau\rangle 
\langle  \partial_X\Psi_0^\sigma,\tilde \psi_n^\sigma\rangle \ |\ X^\tau\big]\\
& = 4 T\Re \mathbb E\big[\langle  \partial_X\Psi_0^\sigma, 
(S_{\tau,\sigma})^*\tilde H^\tau \partial_X\Psi_0^\tau\rangle \ |\ X^\tau\big]\\
& = 4 T\Re \mathbb E\big[\langle  S_{\tau,\sigma}\partial_X\Psi_0^\sigma,
\tilde H^\tau \partial_X\Psi_0^\tau\rangle\ |\ X^\tau\big] .
\end{array}
\end{equation}
In the second step we used that 
\begin{equation}\label{lambda_prim}
\frac{1}{\tilde\lambda_n(X^0)}=
\frac{1}{\tilde\lambda_n(X^\sigma)}(1-\frac{(X^0-X^\sigma)\cdot\partial_X\tilde\lambda_n}{\tilde\lambda_n}).
\end{equation}
where the error term
\begin{equation}\label{gap_variation}
T\sum_{n>0}\frac{(X^0-X^\sigma)\cdot\partial_X\tilde\lambda_n}{\tilde\lambda_n}\le 
|X^0-X^\sigma|_{\ell^\infty} |T\lambda_n^{-1}\lambda_n'|_{\ell^1}= o(M^{-1/2})
\end{equation}
is negligible by the first and last assumption in \eqref{gap_villkor}. %
We see that the main fluctuation term takes the same form as the friction term, so that
the first limit in \eqref{main_lemma_est} proves the second limit and we have obtained the leading
order contribution in the second statement of \eqref{main_lemma_est}. It remains the verify that the other terms in 
\eqref{u_koppling} are negligible, but first comes the proof of Lemma \ref{g_lemma}.

\begin{proof} {\bf (Lemma \ref{g_lemma})}
A way to understand that the large $M^{1/2}$ factor in the $\tilde\psi'$ equation  of  \eqref{X'} does not pollute
the estimate of $G$  in short time intervals is to eliminate $\tilde\psi'$
through  the representation
\[
(\tilde\psi_m)_n'(\tau)=-\sum_k i M^{1/2}\int_0^{ \tau} S_{\tau,\sigma}
(X_k)'_n(\sigma) \partial_{X_k}\tilde H(X^\sigma) S_{\sigma,0}\tilde\psi_m^{ 0}d\sigma \ ,
\]
obtained from the last equation in \eqref{X'}.
The $\tilde\psi'$-term in the $\dot p'$ equation can then
be written as
\[
2\Re\langle \sum_m\tilde\gamma_m(\tilde\psi_m)'_n,\partial_X\tilde H\psi\rangle=
-i2M^{1/2}\Re\langle \int_0^{ \tau} S_{\tau,\varsigma}(X)_n'(\varsigma)\cdot \partial_X\tilde H(X^\varsigma) \psi^\varsigma d\varsigma\ ,\partial_X\tilde H\psi^\tau\rangle,
\]
which by the integration of the first equation in \eqref{X'} and \eqref{koppling} yields the following additional source term in the right hand side to the equation for the Green's function
\begin{equation}\label{S_R_stab}
\begin{split}
R(\tau,\sigma)&:=-i2M^{1/2}\Re\langle \int_\sigma^{ \tau} \int_\sigma^\varsigma S_{\tau,\varsigma}G_{pp}(\varsigma,\upsilon)\cdot \partial_X\tilde H(X^\varsigma)\ \psi^\varsigma,\partial_X\tilde H\psi^\tau\rangle
\ d\upsilon d\varsigma\\
&\le M^{1/2}\int_\sigma^\tau \int_\upsilon^\tau \|G_{pp}(\varsigma,\upsilon)\cdot \partial_X
\tilde H(X^\varsigma)\ \psi^\varsigma\|_{L^2(dx)}\|\partial_X\tilde H\psi^\tau\|_{L^2(dx)} d\varsigma d\upsilon\\
&=M^{1/2}\mathcal O((\tau-\sigma)^2)
\end{split}
\end{equation}
where also the unitary $L^2(dx)$ bound on the solution operator is used.
This remainder leads by Duhamel's representation  to a small contribution 
\[
\int_\sigma^\tau e^{(\tau-\sigma)\tilde A}R(\tau,\sigma)d\sigma= M^{1/2} \mathcal O((\tau-\sigma)^3)
\]
to $G\simeq I+(\tau-\sigma)\tilde A$, 
provided $M^{1/2}(\tau-\sigma)^2=o(1)$, where $\tilde A$ is the submatrix  in \eqref{X'} with $\psi'$ eliminated
\[
\tilde A:= \left[\begin{array}{cc}
0 & \quad\quad I \\
-\lambda_0'' -\langle \psi,\tilde H''\psi\rangle& \quad\quad 0 \\ %
\end{array}\right] .
\]
Therefore, the bound \eqref{f_pp_est} holds for $M^{1/2}(\tau-\sigma)^2=o(1)$
and we will use it for much shorter time intervals, satisfying $M^{1/2}(\tau-\sigma)=\mathcal O(1)$. 

The other coupling in \eqref{u_koppling} based on $\langle \tilde\gamma_m(\tilde\psi_m)_n',\tilde H\Psi_0'\rangle\partial_p u$
with \[
(\tilde\psi_m)_n'(\tau)=  -i\sum_{k=1}^{3N} M^{1/2}\int_0^\tau 
S_{\tau,\sigma} (X_k(\sigma))'_n\cdot\partial_{X_k}\tilde H(X^\sigma)  \tilde\psi_m^\sigma d\sigma\ \]
combined with
\[
(X_k)_n'(\sigma)=\int_0^\sigma \tilde p'(\upsilon)d\upsilon
=\int_0^\sigma\int_0^\upsilon G_{pp}(\upsilon,\varsigma)\langle \tilde\psi_n^\varsigma, 
\tilde H\partial_{X_k}\Psi_0^\varsigma \rangle d\varsigma d\upsilon,
\]
obtained   from \eqref{X'} and \eqref{koppling},
yields by the change of the order of integration and the $L^2(dx)$ bound of $S$ as in \eqref{S_R_stab}
\begin{equation}\label{psi-prim}
\begin{split}
&(\tilde\psi_m)_n'(\tau) =\\
&= -\sum_kiM^{1/2}\int_0^\tau \int_0^\sigma \int_0^\upsilon S_{\tau,\sigma}G_{pp}(\upsilon,\varsigma)\partial_{X_k}
\tilde H^\sigma \langle \tilde\psi_n^\varsigma, \tilde H\partial_{X_k}\Psi_0(X^\varsigma)\rangle 
\tilde\psi_m^\sigma d\varsigma d\upsilon d\sigma \ \\
&=-\sum_k iM^{1/2}\int_0^\tau
\underbrace{\int_\varsigma^\tau \int_\varsigma^\sigma S_{\tau,\sigma}G_{pp}(\upsilon,\varsigma) d\upsilon\  \partial_{X_k}\tilde H(X^\sigma)\tilde\psi_m^\sigma d\sigma}_{L^2(dx)\mbox{-norm}=\mathcal O((\tau-\varsigma)^2)} \langle \tilde\psi_n^\varsigma, \tilde H\partial_{X_k}\Psi_0(X^\varsigma)\rangle d\varsigma\, ,
\end{split}
\end{equation}
which finishes the proof of \eqref{f_pp_est}.
\end{proof}

The quadratic factor $(\tau-\varsigma)^2 d\varsigma$ in \eqref{psi-prim} yields by a change to the fast time scale a factor $M^{-3/2}$,
where $(\tau-\varsigma)^2$ gives the extra  factor  $M^{-1}$ as compared to \eqref{F_uppskatt}, 
and the component wise bound on $\partial_{X_k}\tilde H\psi$ together with the
$\ell^1$ bound on $\tilde H\Psi_0'$ in \eqref{gap_villkor} show that the $\tilde\psi'$-term  in the intrinsic coupling
\eqref{intrinsic} vanishes asymptotically. %

The other terms in \eqref{u_koppling} and \eqref{intrinsic} depending on $X'$ include
the  factor $G_{Xp}(\tau,\sigma)$, which introduces the additional factor $\tau-\sigma$ (as compared to the $p'$  term with the
factor $G_{pp}\sim 1$)  in the slow
time scale, which in the integration of the fast scale $\tau-\sigma=M^{-1/2}\hat \sigma$
yields an extra factor of $M^{-1/2}$ as compared to the case \eqref{fluct_diss} with $G_{pp}(\tau,\sigma)\simeq 1$.
Therefore, the terms depending on $X'$ in \eqref{u_koppling} and \eqref{intrinsic} are asymptotically negligible.

{\it The third limit: the quadratic term.}
The last statement in \eqref{main_lemma_est} 
has by the first condition in \eqref{gap_villkor} and \eqref{gap_variation} its main contribution from the diagonal part
\[
\begin{split}
&\sum_{n=1}^J\mathbb E[|\tilde\gamma_n|^2 \langle \tilde\psi_n^\tau,\partial_X\tilde H(X^\tau)\tilde\psi_n^\tau
\rangle \cdot \partial_p u(X^\tau,p^\tau,\tau)]\\
&=\sum_{n=1}^J 
 \mathbb E[ {T\tilde\lambda_n^{-1}(X^0)}
 \langle \tilde\psi_n^\tau,\partial_X\tilde H(X^\tau)\tilde\psi_n^\tau\rangle
\cdot\partial_p u\big(1+o(1)\big)]\\
&=\sum_{n=1}^J 
 T\, \mathbb E[\langle \tilde\psi_n^\tau,\tilde \lambda^{-1}_n\tilde\lambda_n'(X^\tau) \tilde\psi_n^\tau\rangle
\cdot\partial_p u \big(1+o(1)\big)]\\
&= o(M^{-1/2}).\\
\end{split}
\]
The off diagonal contribution yields as in \eqref{off_diagonal} two factors 
\[
\sum_{n,m} T\int_0^\tau F \tilde\lambda_n^{-1}d\sigma\ T\int_0^\tau F \tilde\lambda_m^{-1}d\sigma
\]
which gives a factor $M^{-1}$ so that the off diagonal part is negligible.

\subsection{The Born-Oppenheimer Approximation}\label{BO_sec} 

The purpose of this section is to study the evolution \eqref{psi_evolution} of $\tilde\psi_n$:
\begin{lemma}\label{born_oppen_lemma}
Assume that $i\dot{\tilde\psi}_n=M^{1/2}\tilde H\tilde\psi_n$ holds with the initial data $\tilde\psi_n^0=\Psi_n(X^0)$,
and that $\tilde H$ satisfies the smoothness estimate \eqref{beta_est},
then the $L^2(dx)$ orthogonal decomposition $\tilde\psi_n=\bar\psi_n \oplus\psi_n^\perp$, where $\bar\psi_n=\alpha\Psi_n$
for some $\alpha\in\mathbb C$,
satisfies 
\begin{equation}\label{omega_estimate}
\langle \psi_n(\tau)^\perp,\psi_n(\tau)^\perp\rangle^{1/2}=\mathcal O(M^{-1/2}).
\end{equation}
\end{lemma}
\begin{proof}

 Let $\psi_n^\tau:=e^{iM^{1/2}\int_0^\tau \tilde\lambda_n^\sigma d\sigma}\tilde\psi_n^\tau$ and make the decomposition
 $\psi_n=\bar\psi_n\oplus\psi_n^\perp$,
where $\bar\psi_n^\tau$ is an eigenvector of $\tilde H(X^\tau)$,  %
satisfying $\tilde H^\tau\bar \psi_n^\tau=\tilde\lambda_n^\tau\bar\psi_n^\tau$
for the eigenvalue $\tilde\lambda_n^\tau\in\rset$. 
The similar decomposition of $\tilde\psi_n$ in the lemma
is related by the factor 
 $e^{iM^{1/2}\int_0^\tau\tilde\lambda_n^\sigma d\sigma}$.
This {\it Ansatz} is motivated by the zero residual
\begin{equation}\label{R_residual}
R\psi_n:= \dot\psi_n +iM^{1/2}(\tilde H-\tilde\lambda_n)\psi_n= 0 
\end{equation}
and the small residual for the eigenvector
\[
\begin{array}{rl}
\langle(\dot{\bar\psi}_n)^\natural,\bar\psi_n\rangle &=0 \\ 
M^{1/2}(\tilde H-\tilde\lambda_n)\bar\psi_n&=0, \end{array}
\]
where 
\begin{equation}\label{natural}
w(X)=\langle \Psi_n(X), w(X)\rangle \Psi_n(X)\oplus w(X)^\natural
\end{equation}
denotes the orthogonal decomposition in the eigenfunction direction $\Psi_n$ and
its orthogonal complement. %
The constructions of the linear operator $R$ in \eqref{R_residual}, 
the orthogonal splitting  $\psi_n=\bar\psi_n\oplus\psi_n^\perp$ and the projection $\natural$ in \eqref{natural} imply
\[
\begin{split}
0 &= \big(R(\bar\psi_n+\psi_n^\bot)\big)^\natural\\
 &= \big(R(\bar\psi_n)\big)^\natural + \big(R(\psi_n^\bot)\big)^\natural\\
 &=  (R\bar\psi_n)^\natural + R(\psi_n^\bot),
 \end{split}
 \] 
 where the last step $\big(R(\psi_n^\bot)\big)^\natural=R(\psi_n^\bot)$ 
 follows from the orthogonal splitting \[\big((\tilde H-\tilde\lambda_n)\psi_n^\bot\big)^\natural=(\tilde H-\tilde\lambda_n)\psi_n^\bot\]
 together with the second order change in the subspace projection
 \[\psi_n^\bot(\tau+\Delta\tau)=\big(\psi_n^\bot(\tau+\Delta \tau)\big)^{\natural(\tau+\Delta \tau)}= 
 \big(\psi_n^\bot(\tau+\Delta \tau)\big)^{\natural(\tau)} + \mathcal O(\Delta \tau^2)
 \]
 which yields $(\dot\psi_n^\bot)^\natural=\dot\psi_n^\bot$; here $\natural(\tau)$ denotes the projection on the orthogonal complement to the eigenvector $\bar\psi_n^\tau$.
   To explain the second order change start with a function $v$ satisfying $\langle v,\Psi_n(X^\tau)\rangle = 0$
  and $\Psi_n(X^\sigma)=\Psi_n(X^\tau) + \mathcal O(\Delta \tau)$ for $\sigma\in [\tau,\tau+\Delta \tau]$
  to obtain
  \[
  \begin{split}
  \big(v^{\natural(\tau+\Delta\tau)}-v^{\natural(\tau)}\big)^{\natural(\sigma)}&= 
 \Big(\langle v,\Psi_n(X^{\tau})\rangle\Psi_n(X^{\tau})
  -  \langle v,\Psi_n(X^{\tau+\Delta\tau})\rangle\Psi_n(X^{\tau+\Delta\tau})\Big)^{\natural(\sigma)}\\
  &= \Big(\mathcal O(\Delta\tau^2) 
  + \langle v,\mathcal O(\Delta\tau)\rangle \Psi_n(X^\tau)\Big)^{\natural(\sigma)}\\
  & =\mathcal O(\Delta\tau^2) 
  + \mathcal O(\Delta\tau)\Big(\Psi_n(X^\tau)-\langle\Psi_n(X^\tau),\Psi_n(X^\sigma)\rangle \Psi_n(X^\sigma)\Big)\\
  &=\mathcal O(\Delta\tau^2).
  \end{split}
  \]

Consequently, the perturbation $\psi_n^\bot$ can be determined from the projected residual
\[
\dot{\psi}_n^\bot =-iM^{1/2}(\tilde H-\tilde\lambda_n)\psi_n^\bot - (R\bar\psi_n)^\natural\]
and we have the solution representation, as in \eqref{sol_repr},
\begin{equation}\label{r_int}
\psi_n^\bot(\tau)= S_{\tau,0}\underbrace{\psi_n^\bot(0)}_{=0}
-\int_0^\tau  S_{\tau,\sigma} \underbrace{\big(R\bar\psi_n(\sigma)\big) ^\natural
}_{=\alpha p\cdot \partial_X\Psi_n}d\sigma,
\end{equation}
where $ S$ is the solution operator $ S_{\tau,0}\psi_n^0=\psi_n^\tau$ in the slow time scale
for the equation 
\[iM^{-1/2}\dot \psi_n^\tau=\big(\tilde H(X^\tau)-\tilde\lambda_n(X^\tau)\big)\psi_n^\tau.\]
Integration by parts introduces the factor $M^{-1/2}$ we want
\begin{equation}\label{s_int_def}
\begin{split}
\int_0^\tau S_{\tau,\sigma} R\bar\psi_n(\sigma) ^\natural d\sigma&=
\int_0^\tau iM^{-1/2}\frac{d}{d\sigma} (S_{\tau,\sigma})(\tilde H-\tilde\lambda_n)^{-1}  R\bar\psi_n(\sigma) ^\natural d\sigma\\
&=\int_0^\tau iM^{-1/2}\frac{d}{d\sigma} \big(S_{\tau,\sigma}(\tilde H-\tilde\lambda_n)^{-1}  
R\bar\psi_n(\sigma) ^\natural\big) d\sigma\\
&\qquad -\int_0^\tau iM^{-1/2} S_{\tau,\sigma}\frac{d}{d\sigma}
\big((\tilde H-\tilde\lambda_n)^{-1}(X^\sigma)  R\bar\psi_n(\sigma) ^\natural\big) d\sigma \\
&= \frac{i}{M^{1/2}} (\tilde H-\tilde\lambda_n)^{-1}  R\bar\psi_n(\tau) ^\natural
- \frac{i}{M^{1/2}} S_{\tau,0}(\tilde H-\tilde\lambda_n)^{-1}  R\bar\psi_n(0) ^\natural\\
&\qquad -\int_0^\tau iM^{-1/2} S_{\tau,\sigma}\frac{d}{d\sigma}\big((\tilde H-\tilde\lambda_n)^{-1}(X^\sigma)  
R\bar\psi_n(\sigma) ^\natural\big) d\sigma.\\
\end{split}
\end{equation}
The spectral gap assumption in \eqref{gap_villkor},
i.e.  $|\tilde\lambda_m-\tilde\lambda_n|>c$ for $m\ne n$
(which excludes multiple eigenvalues), implies by diagonalization on the orthogonal complement of $\Psi_n$
\begin{equation}\label{vtilde}
\begin{split}
\|(\tilde H-\tilde\lambda_n)^{-1}  R\bar\psi_n(0) ^\natural\|_{L^2(dx)}^2 &=
\sum_{m\ne n}^\infty (\tilde\lambda_m-\tilde\lambda_n)^{-2}|\langle \Psi_m,R\bar\psi_n(0) ^\natural\rangle|^2\\
 &=\sum_{m\ne n}^\infty (\tilde\lambda_m-\tilde\lambda_n)^{-2}\tilde\lambda_m^{-1}
|\langle \Psi_m,\tilde H^{1/2}R\bar\psi_n(0) ^\natural\rangle|^2\\
&\le c^{-3}\|\tilde H^{1/2}R\bar\psi_n(0) ^\natural\|_{L^2(dx)}^2=\mathcal O(c^{-3})
\end{split}
\end{equation}
and  the analogous estimate
\[
\|\frac{d}{d\sigma}\big((\tilde H-\tilde\lambda_n)^{-1}(X^\sigma)  R\bar\psi_n(\sigma) ^\natural\big)\|_{L^2(dx)}
=\mathcal O(c^{-5/2}),
\]
inserted  in \eqref{r_int} proves the Lemma
for bounded time intervals.

The evolution on longer times requires another idea:
one can integrate by parts recursively in \eqref{s_int_def}  to obtain the expansion
\[
\begin{split}
\int_0^\tau S_{\tau,\sigma} R\bar\psi_n(\sigma) ^\natural d\sigma
&=\Big[S_{\tau,\sigma}\Big(\beta \tilde R -\beta \frac{d}{d\sigma}(\beta\tilde R) + \beta \frac{d}{d\sigma}
\big(\beta \frac{d}{d\sigma}(\beta\tilde R)\big)
-\ldots\Big)\Big]_{\sigma=0}^{\sigma=\tau}\, ,\\
\beta &:= iM^{-1/2}(\tilde H-\tilde\lambda_n)^{-1},\\
\tilde R &:= R\bar\psi_n(\sigma) ^\natural,
\end{split} \]
 so that by \eqref{r_int} we have
  \[
      \psi_n^\bot(\tau)= S_{\tau,0} \psi_n^\bot(0)
     - \left[S_{\tau,\sigma}\Big(\beta \tilde R -\beta \frac{d}{d\sigma}(\beta\tilde R) + \beta \frac{d}{d\sigma}
               \big(\beta \frac{d}{d\sigma}(\beta\tilde R)\big)-\ldots\Big)\right]_{\sigma=0}^{\sigma=\tau}\, .
\]
 By choosing 
  
  \[
  \bar\psi_n^\bot(\sigma)\Big|_{\sigma=0}
  =-\Big(\beta \tilde R(\sigma) -\beta \frac{d}{d\sigma}(\beta\tilde R)(\sigma) + \beta \frac{d}{d\sigma}
               \big(\beta \frac{d}{d\sigma}(\beta\tilde R)\big)(\sigma)-\ldots\Big)\Big|_{\sigma=0},
\] 
  we  get
  \begin{equation}\label{beta_exp}
  \bar\psi_n^\bot(\tau) =  -\sum_{k=0}^\infty \beta_0^k R_0(\tau)\, ,
  \end{equation}
  where $\beta_0:=-iM^{-1/2}(\tilde H-\tilde\lambda_n)^{-1} \frac{d}{d\tau}$ and 
  $R_0:=iM^{-1/2}(\tilde H-\tilde\lambda_n)^{-1}\tilde R$.
  We assume this expansion \eqref{beta_exp} is convergent in $L^2(dx)$ for each $\tau$, which follows from the smoothness  assumption
\begin{equation}\label{beta_est}
\mbox{$\|\beta_0^kR_0(\tau)\|_{L^2(dx)}\rightarrow 0$ as $k\rightarrow\infty$}
\end{equation}
 and \eqref{vtilde}.

\end{proof}

\subsection{Proof of Theorem \ref{long_time_thm}}
\begin{proof}
Use the estimate
\[
\begin{split}
&\mathcal T^{-1}\int_0^{\mathcal T}
\mathbb{E}_\gamma\big[  g(X^{\tau},p^{\tau})\big]
-\mathbb E_W[g( X_L^\tau, \dot{X}_L^\tau)\big] \ d\tau\\
&=o(M^{-1/2}) \mathcal T^{-1}\int_0^{\mathcal T} \int_0^\tau \mathbb E_\gamma[
|\mathcal D u(X^\sigma,p^\sigma,\sigma;\tau)|_{\ell^1}] d\sigma d\tau,
\end{split}
\]
from \eqref{main_estimate} and \eqref{u_koppling} in the proof of Theorem \ref{thm},
to obtain
\[
\begin{split}
&\mathcal T^{-1}\int_0^{\mathcal T}
\mathbb{E}_\gamma\big[  g(X^{\tau},p^{\tau})\big]
-\mathbb E_W[g( X_L^\tau, \dot{X}_L^\tau)\big] \ d\tau
=o(1),
\end{split}
\]
based on the assumption 
\[
\int_0^{\tau} |\mathcal Du(X^\sigma,p^\sigma,\sigma;\tau)|_{\ell^1} d\sigma=\mathcal O(M^{1/2}),
\]
where $\mathcal D u$ is used for the combination of derivatives $\partial_p u, \partial_{pp} u,\partial_{Xp}u$
appearing in the estimates of the proof of Theorem \ref{thm}.
\end{proof}

\subsection{A Motivation for $\lim_{\tau\rightarrow\infty}\int_0^\tau 
\partial_p u(X^\sigma,p^\sigma,\sigma;\tau) d\sigma=\mathcal O(M^{1/2})$}\label{magnus_sec}
This section verifies the assumption in Theorem \ref{long_time_thm} that
 $\int_0^\infty \partial_p u(X^\sigma,p^\sigma,\sigma) d\sigma=\mathcal O(M^{1/2})$, in the
 special case that the friction matrix is a constant positive multiple of the identity matrix. %

Using the expected value $\mathbb E=\mathbb E_W$ with respect to 
the Wiener process in this section, we have the representation,
\begin{equation}\label{u_p_repr}
\begin{split}
\partial_p u(X,p,\sigma;\tau) &=\partial_p \mathbb E[ g(X_L^\tau, p_L^\tau)\ |\  X_L^\sigma=X, p_L^\sigma=p]\\
&=  \mathbb E[ \partial_{p_L^\sigma} g(X_L^\tau, p_L^\tau)\ |\  X_L^\sigma =X, p_L^\sigma =p]\\
&=  \mathbb E[ \partial_{X} g(X_L^\tau, p_L^\tau) 
\frac{\partial X_L^\tau}{\partial p_L^\sigma}
+ \partial_{p} g(X_L^\tau, p_L^\tau) \frac{\partial p_L^\tau}{\partial p_L^\sigma}
\ |\  X_L^\sigma=X, p_L^\sigma=p]
\end{split}
\end{equation}
where the stochastic flow $(\frac{\partial X_L^\varsigma}{\partial p_L^\sigma}, \frac{\partial p_L^\varsigma}{\partial p_L^\sigma})=: \big(X_L'(\varsigma;\sigma),p_L'(\varsigma;\sigma)\big)$
solves the linear equation 
\[
\begin{split}
\frac{d}{d\varsigma} X_L'(\varsigma;\sigma) &= p_L'(\varsigma;\sigma)\\
\frac{d}{d\varsigma} p'_L(\varsigma;\sigma) &= -\partial_{XX}\lambda_0(X_L^\varsigma)X_L'(\varsigma;\sigma) -\bar K p_L'(\varsigma;\sigma)
\end{split}
\]
in the special case when diffusion coefficient $\bar K:=M^{-1/2}K=k\, I$ is a constant multiple of the identity,
with $k>0$. To simplify the writing we use the notation $\partial_{XX}\lambda_0(X)=:\lambda_0''(X)$ for the
Hessian of $\lambda_0$.
Let the %
matrix in the right hand side be denoted by
\[
\hat A(\varsigma):= \left[\begin{array}{cc}
0 & \quad \quad I \\
-\lambda_0''(X_L^\varsigma) & \quad\quad -\bar K \\ 
\end{array}\right] \ .
\]
The discrete time levels $\sigma=\varsigma_0, \ldots, \varsigma_{\hat N}=\tau$ yields for sufficiently small time step $\Delta t:=\varsigma_{n+1}-\varsigma_n$,
the representation
\begin{equation}\label{x_p_repr}
\left[\begin{array}{c}
X_L'(\varsigma;\sigma)\\
p'_L(\varsigma;\sigma)\\
\end{array}\right]
\simeq \big (\prod_{n=1}^{\hat N} e^{\Delta t \hat A(\varsigma_n)}\big) \left[\begin{array}{c}
0\\
I\\
\end{array}\right]
:= e^{\Delta t \hat A(\varsigma_N)} e^{\Delta t \hat A(\varsigma_{N-1})} \ldots e^{\Delta t \hat A(\varsigma_1)}\left[\begin{array}{c}
0\\
I\\
\end{array}\right]
\, .
\end{equation}
To estimate this product we will use 
the Euclidian matrix norm \[\| A\|:=\sup_{Y\in\rset^{6N}} |AY|/|Y|,\] which  is bounded by the
square root of the largest eigenvalue of $A^*A$ and satisfies the product rule
\[
\|\prod_{n=1}^{\hat N} e^{\Delta t \hat A(\varsigma_n)}\|\le \prod_{n=1}^{\hat N}\| e^{\Delta t \hat A(\varsigma_n)}\|.
\]

To study these exponentials, we need information about the spectrum and therefore  we diagonalize the matrix $\hat A(\varsigma_n)$
and define the matrix
\[
Q= \left[\begin{array}{cc}
a_+ & \quad \quad a_- \\
I & \quad\quad I \\ 
\end{array}\right] ,
\]
where
\[
a_\pm := -\frac{1}{2} \bar K \pm i \big(\lambda_0''(\varsigma_n) - 4^{-1}\bar K^2\big)^{1/2}.
\]
The matrix $Q$ transforms $\hat A(\varsigma_n)$ into block diagonal form
\[
Q^{-1}\hat A(\varsigma_n)Q 
=\left[\begin{array}{cc}
a_+ & \quad \quad 0 \\
0 & \quad\quad a_- \\ 
\end{array}\right] ,
\]
since $\bar K$ and $\lambda''_0$ commute (it is only here we use that $\bar K $ is a multiple of the identity).
Write the eigenvalues of the Hermitian matrices $a_\pm$ as $-k/2\pm i (\bar\lambda_m - k^2/4)^{1/2}$.
When $\bar\lambda_m - k^2/4$ is positive, the real part of the eigenvalue is negative equal to $-k/2$
and when $\bar\lambda_m - k^2/4$ is negative the real part of the eigenvalue is bounded by $(k^2/4-\bar\lambda_m)^{1/2} - k/2$.
Introduce therefore the function 
\[
-k/2 + (k^2/4-\bar\lambda(\varsigma_n))_+^{1/2}:= -k/2 + \sqrt{\max_m(0,k^2/4-\bar\lambda_m(\varsigma_n))},
\]
which bounds the real part of the eigenvalue,
to obtain the spectral bound
\[
\|e^{\Delta t \hat A(\varsigma_n)}\| \le \exp\Big(\Delta t\big( 
-k/2 + (k^2/4-\bar\lambda(\varsigma_n))_+^{1/2}
\big)\Big)
\]
and consequently
\begin{equation}\label{exp_k}
\begin{split}
\|\prod_{n=1}^{\hat N} e^{\Delta t \hat A(\varsigma_n)}\|  &\le
\prod_{n=1}^{\hat N}\exp\Big(\Delta t\big( 
-k/2 + (k^2/4-\bar\lambda(\varsigma_n))_+^{1/2}
\big)\Big)\\
&=\exp\Big(\sum_{n=1}^{\hat N}\Delta t\big( 
-k/2 + (k^2/4-\bar\lambda(\varsigma_n))_+^{1/2}
\big)\Big)\\
&\simeq
\exp\Big(\int^\tau_\sigma \big( 
-k/2 + (k^2/4-\bar\lambda(\varsigma))_+^{1/2}
\big)d\varsigma\Big).
\end{split}
\end{equation}
The theory of large deviations tells us that for low temperature $T\ll 1$, Langevin solution paths $X_L^\varsigma$ spend
long time around stable equilibria, where $\bar\lambda_m>0$, and at some rare events they
make short time $\tau_e$ (of order one in the slow scale) excursions  between such equilibria, see Ref.   \cite{freidlin}.
The number of such rare events in a time interval $[0,\tau-\sigma]$ can be approximately modelled by a Poisson
process $m_{\tau-\sigma}$ with the intensity $\xi$, proportional to $e^{\Delta \lambda_0/T}\sim e^{-1/T}$
(for a negative potential difference $-\Delta\lambda_0:=\lambda_0(X)-\lambda_0(Y)\sim 1$).
Let $\kappa:=\max_{X}\big(k^2/4-\bar\lambda(X)\big)_+^{1/2}$,  and $\beta:=t_e\kappa$.

The estimates \eqref{x_p_repr}  and \eqref{exp_k} together show that
\[
\begin{split}
\Big | \left[\begin{array}{c}
\partial_{p_k}X_L(\varsigma;\sigma)\\
\partial_{p_k}p_L(\varsigma;\sigma)\\
\end{array}\right]\Big | 
&\le 
\exp\Big(\int^\tau_\sigma \big( 
-k/2 + (k^2/4-\bar\lambda(\varsigma))_+^{1/2}
\big)d\varsigma\Big)
\end{split}
\]
so that representation  \eqref{u_p_repr}  implies the bound 
\[
\begin{split}
\lim_{\mathcal T\rightarrow\infty}
&\mathcal T^{-1}\int_0^\mathcal T \int_0^\tau |\partial_p u(X^\sigma,p^\sigma,\sigma;\tau)|_{\ell^1} d\sigma d\tau\\
&\qquad \le C  \lim_{\mathcal T\rightarrow\infty} \mathcal T^{-1}\int_0^\mathcal T \int_0^\tau
e^{\int^\tau_\sigma -k/2 +(k^2/4-\bar\lambda(\varsigma))_+^{1/2} d\varsigma} d\sigma d\tau\\
&\qquad =C \lim_{\tau\rightarrow\infty} \int_0^\tau
\mathbb E[e^{\int_\sigma^\tau -k/2 +(k^2/4-\bar\lambda(\varsigma))_+^{1/2} d\varsigma}] d\sigma 
\end{split}
\]
for some constant $C$, where the expected value is taken with respect to the Poisson process.
This expected value can then roughly by estimated by
\[
\begin{split}
\mathbb E[e^{\int_\sigma^\tau -k/2 +(k^2/4-\bar\lambda(\varsigma))_+^{1/2} d\varsigma}] &\le 
\mathbb E[e^{-k(\tau-\sigma)/2 +\beta m_{\tau-\sigma}}]\\
&=e^{-(\xi+k/2)(\tau-\sigma)}\sum_{m=0}^\infty e^{\beta m}\frac{\big(\xi(\tau-\sigma)\big)^m}{m!}\\
&=e^{\big((e^{\beta}-1)\xi-k/2\big)(\tau-\sigma)}.
\end{split}
\]
Since $T\ll \log M$, we have a tiny intensity $\xi\sim e^{-1/T}\ll k\sim M^{-1/2}$, which implies
 $(e^{\beta}-1)\xi-k/2<-k/3$ and we conclude that 
 \[
\lim_{\mathcal T\rightarrow\infty} \mathcal T^{-1}\int_0^\mathcal T \int_0^\tau 
|\partial_p u(X^\sigma,p^\sigma,\sigma;\tau)|_{\ell^1} d\sigma d\tau =\mathcal O(M^{1/2}).
 \]

\subsection{A stochastic limit in \eqref{K_definition}?}\label{limit_K}
A conjecture to obtain a  stochastic continuum limit in \eqref{K_definition} based on 
random matrices can be formulated as follows. 
 We may assume that $\tilde H(X^{\tau-M^{-1/2}\hat\sigma})$ does not converge 
 but its average along the solution path becomes
 random in the sense 
 that the solution operator $\tilde S_{\hat\tau,\hat\tau-\hat\sigma}$  tends to $e^{-i\hat\sigma\tilde H_\infty}$,
 where $\tilde H_\infty$ is a random matrix, as $M\rightarrow\infty$. %
 The motivation for this guess is that 
 the solution operator $\tilde S_{\hat\tau,\hat\tau-\hat\sigma}$ for small $\hat\sigma$ to %
 leading order is
$ e^{-i\int^{\hat\sigma}_{0} \tilde H(X^{\tau-M^{-1/2}\hat\varsigma})d\hat\varsigma} %
$, in a Magnus expansion \cite{magnus}, and %
a tiny time step of size
$M^{-1/2}$ makes an increment  $M^{-1/2}p$ in the position $X$, which yields an increment of size $\mathcal O(M^{-1/2})$ in each matrix component $\tilde H_{ij}$. If these increments in the matrix components become random,
due to sensitivity and fluctuation of $X$,   and the size $J$ of the matrix $\tilde H$ grows as $M$ tends to infinity, the fluctuations in the eigenvalues of the matrix could lead to a non trivial stochastic limit of the spectrum of the solution operator as $M$ tends to infinity, analogously to what happens in random matrix theory.  An example of a random matrix result, which gives an indication of the required size of $J$ compared to $M$, is
Wigners Semi-Circle Law saying that symmetric matrices of size $J$, with independent  identically distributed entries
with mean zero, variance $J^{-1}$ and bounded moments, %
have the density of eigenvalues 
semi-circle distributed asymptotically as $J\rightarrow\infty$, i.e. the continuum density is
$\sqrt{4-\rho^2}1_{|\rho|\le 2} \, d\rho/(2\pi)$, cf. Ref.   \cite{random}.
Since in our case the variance of the matrix elements could be $\mathcal O(M^{-1})$,
the choice  $J\sim M$ might fill spectral gaps of size one, if the matrix elements were independent identically distributed.
On the other hand, we do not know the distribution of our matrix elements (in particular they are not independent), so
the existence of a stochastic limit giving a continuum eigenvalue distribution is a vague idea.

Assuming the existence of the following spectral representation, the limit \eqref{K_definition}
can by the Fourier transform, $\mathcal{F}h(\hat\sigma):=\int_\rset e^{i\hat\sigma\lambda}h(\lambda)d\lambda$ and its inverse $h(\lambda)=2\pi \int_\rset e^{-i\hat\sigma\lambda}\mathcal F h(\hat\sigma)d\hat\sigma$, 
be written
\begin{equation}\label{fourier_representation}
\begin{split}
K(X^\tau) &=
\int_0^\infty \underbrace{\int_0^\infty 2\cos (\hat \sigma\lambda) \Gamma(\lambda,X^\tau)d\lambda}_{=\mathcal{F}\, \Gamma (\hat\sigma,X^\tau)}\ d\hat \sigma\\
&=\int_0^\infty \mathcal{F} \, \Gamma(\hat\sigma,X^\tau) d\hat \sigma\\
&= \pi\,  \Gamma (0,X^\tau),
\end{split}
\end{equation}
where  $\Gamma \in L^1(\rset)\cap L^2(\rset)\cap \mathcal C(\rset)$ is defined by  a %
spectral decomposition, based on the eigenfunctions $\{\Psi_j^\infty\}_{j=1}^\infty$ and eigenvalues 
$\{\tilde\lambda_j\}_{j=1}^\infty$ of $\tilde H_\infty$,
\[
\begin{split}
&\mathbb E[ \langle   \cos(\hat \sigma\tilde H_\infty) \partial_{X_n}\Psi_0(X),\tilde H_\infty(X)\partial_{X_m}
\Psi_0(X)\rangle \ | \ X] \\
&=
\sum_{j=1}^\infty\mathbb E[ \cos(\hat \sigma \tilde\lambda_j)  \langle \partial_{X_n}\Psi_0(X), \Psi_j^\infty\rangle 
\langle \tilde H_\infty\partial_{X_m}\Psi_0(X), \Psi_j^\infty\rangle  \ | \ X] \\
&=
\int_\rset \cos(\hat \sigma \lambda) 
\underbrace{
\sum_{j\in\{n:\, \tilde\lambda_n\in [\lambda,\lambda+d\lambda)\} }
{\mathbb E[ \langle \partial_{X_n}\Psi_0, \Psi_j^\infty\rangle 
\langle \tilde H_\infty\partial_{X_m}\Psi_0, \Psi_j^\infty\rangle  \ | \ X]} %
}_{=:\Gamma _{nm} (\lambda,X)d\lambda},%
\end{split}
\]
and the expected value is with respect to the random matrix $\tilde H_\infty$.

For assumption \eqref{K_definition} to make sense
the eigenvalue distribution must approach a continuum density, so that
the integral kernel decays in time  and  
$ \Gamma (0,X)$ is well defined; to have a positive limit
it is necessary that the density of states does not vanish around the origin, which excludes
a spectral gap in $\tilde H_\infty$ around the origin.
 The condition
$\Gamma (0,X)$ being positive semi-definite  can also imply that 
$\langle \Psi_0',\Psi_0'\rangle$ 
is infinite -- Section \ref{fric_sec} gives a motivation for such a setting based on the Thomas-Fermi model.

\subsection{A Motivation for Non Zero Friction}\label{fric_sec}
Note that to have non zero friction and dissipation requires 
$\Gamma (0,X)>0$,
in \eqref{fourier_representation},
which together with
$\langle \tilde H^{1/2}\Psi_0',\tilde H^{1/2}\Psi_0'\rangle\simeq \int_\rset \Gamma (\lambda,X)d\lambda$ being finite
implies that %
$\langle \Psi_0',\Psi_0'\rangle\simeq \int_\rset \lambda^{-1}\Gamma (\lambda,X)d\lambda=\infty$.
To have an infinite $L^2(dx)$-norm is possible with a slow decay as $|x_j|\rightarrow\infty$, for instance
that $\partial_{X_n} \Psi_0$ decays as $|x_j-X_n|^{-\alpha}$ with $1/2<\alpha\le 3/2$ for large $|x_j| \ j=1,\ldots J\, ,$
using spherical symmetry.
Is such a decay  reasonable in reality? The Thomas-Fermi model can be used
to motivate that $\partial_X\Psi_0$ could decay as $|x|^{-1}$.

The {\it Thomas-Fermi model} for ground state energies, using an electron density in $\rset^3$,
is asymptotically exact  for large systems where the number of electrons, $J$, and the total nuclear charge, $Z$, tend to
infinity, see Ref.   \cite{lieb}. Its electron density 
\begin{equation}\label{rho_defin}
\rho=\Phi^{3/2}/\gamma, \qquad \gamma:=(3\pi^2)^{2/3},
\end{equation}
satisfies
for the neutral case $J=Z$ the Thomas-Fermi equation
\[
-\Delta_x \Phi + \frac{4}{3\pi}\Phi^{3/2}=4\pi\sum_{n=1}^N Z_n \delta( x-X_n),
\]
and for the derivative
\begin{equation}\label{rho_deriv}
\partial_X \rho=3\Phi^{1/2}\partial_X\Phi/(2\gamma),
\end{equation} 
we obtain
\[
-\Delta_x \partial_{X_n}\Phi + \frac{2}{\pi}\Phi^{1/2}\partial_{X_n}\Phi= -4\pi Z_n \delta'( x-X_n).
\]
The solution $\partial_X\Phi$ then roughly  behaves as the derivative of the Green's function,
which decays as $|x|^{-2}$. One can observe the decay
$|\Psi_0|=\rho^{1/2}\sim |x|^{-3}$, see Ref.   \cite{lieb}. This implies by \eqref{rho_deriv} and \eqref{rho_defin}
that $\partial_X\Psi_0\sim\partial_X\rho^{1/2}=\partial_X\rho/(2\rho^{1/2}) $ has the desired decay
\[
\partial_X\rho/\rho^{1/2}\sim\Phi^{1/2}\partial_X\Phi/\Phi^{3/4}\sim\partial_X\Phi/\Phi^{1/4}\sim |x|\partial_X\Phi \sim |x|^{-1}.
\]

\section*{Acknowledgment}
This  work is supported by the Swedish Research Council grant 621-2010-5647.

\end{document}